\documentclass[aip,graphicx,amsmath,amssymb]{revtex4-1}

\usepackage{graphicx}
\graphicspath{{figures/}}
\usepackage{bm}
\usepackage[english]{babel}
\makeatletter
\@namedef{l@en}{\l@english}
\makeatother

\newcommand{\ga}{$\text{Ga}_2\text{O}_3$}

\newcommand{\meVA}{meV/$\text{\AA}$}

\draft 

\begin{document}


\title{A Neuroevolution Potential for Gallium Oxide: \\
Accurate and Efficient Modeling of Polymorphism and Swift Heavy-Ion Irradiation}



\author{Yaohui Gu}
\affiliation{State Key Laboratory of Heavy Ion Science and Technology, Institute of Modern Physics, Chinese Academy of Sciences, Lanzhou 730000, China}
\affiliation{School of Nuclear Science and Technology, University of Chinese Academy of Sciences, Beijing 100049, China}
\author{Binbo Li}
\thanks{Y.G. and B.L. contributed equally to this work.}
\affiliation{State Key Laboratory of Heavy Ion Science and Technology, Institute of Modern Physics, Chinese Academy of Sciences, Lanzhou 730000, China}
\affiliation{School of Nuclear Science and Technology, University of Chinese Academy of Sciences, Beijing 100049, China}
\author{Linyang Jiang}
\affiliation{State Key Laboratory of Heavy Ion Science and Technology, Institute of Modern Physics, Chinese Academy of Sciences, Lanzhou 730000, China}
\affiliation{School of Nuclear Science and Technology, University of Chinese Academy of Sciences, Beijing 100049, China}
\author{Yuhui Hu}
\affiliation{State Key Laboratory of Heavy Ion Science and Technology, Institute of Modern Physics, Chinese Academy of Sciences, Lanzhou 730000, China}
\affiliation{School of Nuclear Science and Technology, University of Chinese Academy of Sciences, Beijing 100049, China}
\affiliation{School of Materials $\&$ Energy, Lanzhou University, Lanzhou, 730000, China}
\author{Wenqiang Liu}
\affiliation{State Key Laboratory of Heavy Ion Science and Technology, Institute of Modern Physics, Chinese Academy of Sciences, Lanzhou 730000, China}
\affiliation{School of Nuclear Science and Technology, University of Chinese Academy of Sciences, Beijing 100049, China}
\author{Lijun Xu}
\affiliation{State Key Laboratory of Heavy Ion Science and Technology, Institute of Modern Physics, Chinese Academy of Sciences, Lanzhou 730000, China}
\affiliation{School of Nuclear Science and Technology, University of Chinese Academy of Sciences, Beijing 100049, China}
\author{Pengfei Zhai}
\affiliation{State Key Laboratory of Heavy Ion Science and Technology, Institute of Modern Physics, Chinese Academy of Sciences, Lanzhou 730000, China}
\affiliation{School of Nuclear Science and Technology, University of Chinese Academy of Sciences, Beijing 100049, China}
\author{Haizhou Xue}
\affiliation{State Key Laboratory of Heavy Ion Science and Technology, Institute of Modern Physics, Chinese Academy of Sciences, Lanzhou 730000, China}
\affiliation{School of Nuclear Science and Technology, University of Chinese Academy of Sciences, Beijing 100049, China}
\author{Jie Liu}
\affiliation{State Key Laboratory of Heavy Ion Science and Technology, Institute of Modern Physics, Chinese Academy of Sciences, Lanzhou 730000, China}
\affiliation{School of Nuclear Science and Technology, University of Chinese Academy of Sciences, Beijing 100049, China}
\author{Jinglai Duan}
\thanks{Corresponding author. Email: j.duan@impcas.ac.cn}
\affiliation{State Key Laboratory of Heavy Ion Science and Technology, Institute of Modern Physics, Chinese Academy of Sciences, Lanzhou 730000, China}
\affiliation{School of Nuclear Science and Technology, University of Chinese Academy of Sciences, Beijing 100049, China}
\affiliation{Advanced Energy Science and Technology Guangdong Laboratory, Huizhou 516000, China}


\date{\today}

\begin{abstract}
Gallium oxide (\ga) is a wide-bandgap semiconductor with promising applications in 
high-power and high-frequency electronics. 
However, its complex polymorphic nature poses substantial challenges for fundamental 
studies, particularly in understanding phase-transformation behaviors under nonequilibrium conditions.
Here, we develop a robust, accurate, and computationally efficient machine-learning 
interatomic potential (MLIP) for \ga\ based on the neuroevolution potential (NEP) 
framework combined with an energy-dependent weighting strategy. 
The resulting NEP potential demonstrates clear advantages over the 
state-of-the-art tabGAP potential with respect to both accuracy and computational efficiency. 
Furthermore, we introduce a physically process-oriented sampling strategy to 
systematically augment the training dataset, thereby enhancing the MLIP performance 
for targeted physical phenomena. 
As a representative application, a dedicated NEP potential is constructed for swift heavy-ion (SHI) 
irradiation simulations of $\beta$-\ga. 
The simulated results are in quantitative agreement with experimental 
observations and provide a consistent physical explanation for the reported 
experimental discrepancies regarding phase transformations in the ion track of $\beta$-\ga.
\end{abstract}


\maketitle 

\section{Introduction}

Gallium oxide (\ga), a representative of the fourth generation of 
semiconductors, has garnered widespread interest due to its unique physical and 
electronic properties.\cite{pearton_review_2017} 
With an ultra-wide bandgap, high breakdown electric field, and transparency in the 
deep ultraviolet region, \ga\ holds significant promise for applications including power 
electronics,\cite{tadjer_toward_2022} solar-blind UV photodetectors\cite{pratiyush2019advances} 
and gas sensors.\cite{zhu_gallium_2022} 
In addition to its thermodynamically stable monoclinic $\beta$ phase, \ga\ exhibits a 
rich polymorphism, including metastable $\alpha$ (corundum-type), $\gamma$ 
(defective spinel), $\delta$ (bixbyite), and $\kappa$ (orthorhombic) phases.\cite{yoshioka_structures_2007} 
This polymorphic diversity opens new avenues for directional material design and property 
tuning but also presents formidable challenges in controlled synthesis, phase stability, 
and interface engineering.
Consequently, a comprehensive theoretical understanding of \ga---particularly of its 
complex polymorphic nature---is essential for realizing its full potential in practical 
applications.

Early theoretical investigations of \ga\ focused on its electronic structure 
using first-principles methods,\cite{he_first-principles_2006,xu_structure_2007,yoshioka_structures_2007} 
laying the groundwork for understanding its doping behavior,\cite{varley_oxygen_2010} 
lattice dynamics,\cite{yan_phonon_2018,yang_lattice_2024}
interface states,\cite{bermudez_structure_2006} and phase stability.\cite{fan_low-energy_2022} 
While first-principles approaches continue to offer critical insights, their high computational 
cost and limited scalability pose significant challenges for modeling large-scale phenomena, 
such as irradiation-induced phase transitions or extended defect evolution.\cite{azarov_disorder-induced_2021,han_unraveling_2025} 
Conversely, molecular dynamics (MD) enables simulations at larger time 
and length scales; however, the empirical interatomic potentials 
it relies on often struggle to accurately capture the complex potential energy landscape of 
\ga's polymorphic phases.\cite{blanco_energetics_2005} 

In recent years, the rapid advancement of artificial intelligence has ushered in a 
paradigm shift in materials science.\cite{pyzer-knapp_accelerating_2022} 
Among the most transformative developments is the emergence of machine learning interatomic 
potentials (MLIPs), which offer near-DFT accuracy combined with excellent scalability, 
making them powerful alternatives to traditional empirical potentials 
in MD simulations.\cite{ko_recent_2023} 
Motivated by these advantages, several pioneering research groups have made 
continuous efforts to develop MLIPs tailored for \ga, spanning a wide range of polymorphic 
phases---from the thermodynamically stable $\beta$ phase\cite{li_deep_2020,liu_machine_2020,zhao_phase_2021} 
to metastable $\alpha$,\cite{sun_neuroevolution_2023} $\kappa$,\cite{wang_dissimilar_2024} 
$\epsilon$,\cite{sun_neuroevolution_2023} and amorphous configurations.\cite{liu_unraveling_2023} 
A landmark achievement was realized by Zhao \textit{et al.},\cite{zhao_complex_2023} 
who introduced a general-purpose MLIP referred to as tabGAP, aiming to accurately 
describe the atomic interactions across all major \ga\ polymorphs. 
This potential has enabled large-scale simulations of 
defect evolution,\cite{he_threshold_2024,he_ultrahigh_2024,liu_orientation-dependent_2025} 
phase transformations\cite{azarov_self-assembling_2025} 
and crystallization dynamics,\cite{zhang_orientation-dependent_2023,li_edge-dependent_2025} 
and has played a pivotal role in recent studies of irradiation effects in 
\ga.\cite{azarov_universal_2023,zhao_crystallization_2025,han_electronic_2025,abdullaev_ions_2025} 
However, despite its considerable success, tabGAP still exhibits notable limitations compared 
to other state-of-the-art MLIPs, including relatively lower accuracy and computational 
speed, as will be demonstrated in the following sections. 
These limitations motivate us to rigorously benchmark the overall 
performance of tabGAP and develop a new, more accurate and efficient MLIP for \ga.

In this work, we adopt the neuroevolution potential (NEP) framework, which is well known
for its high computational efficiency,\cite{xu_gpumd_2025} to construct MLIPs for \ga.
To capture the complex potential-energy landscape of Ga--O compounds, we propose an
energy-dependent weighting strategy for training.
With this strategy, the resulting NEP model achieves consistently higher prediction
accuracy than tabGAP across all energy ranges and for most configuration types.
Furthermore, we find that both MLIP architectures exhibit varying degrees of zero-shot
prediction limitations as well as systematic softening of the energy landscape.
To mitigate these issues, we further augment the training dataset with additional
configurations generated following physically motivated sampling rules.
As a representative application, to clarify recent controversies surrounding swift heavy-ion (SHI) 
irradiation experiments on $\beta$-\ga, we train a dedicated NEP model using a
dataset augmented with additional $\gamma$-phase configurations and $\beta$-phase
heating--cooling pathways.
High-throughput MD simulations based on this model yield a series of irradiated
$\beta$-\ga\ structures across a wide range of electronic energy losses.
The simulated results are in quantitative agreement with experimental observations and
provide a consistent physical explanation for the reported experimental discrepancies.

\section{Theory}

\subsection{NEP architecture}

NEP, developed by Fan \textit{et al.}, consists of an ACE-like descriptor constructed using 
Chebyshev polynomials and a regression model based on 
a feedforward neural network.\cite{fan_neuroevolution_2021,fan_gpumd_2022,song_general-purpose_2024} 
Its explicit atomic-environment featurization is carefully designed to strike an excellent 
balance between prediction accuracy and computational efficiency. 
Moreover, NEP is natively implemented in the Graphics Processing Units Molecular Dynamics 
(GPUMD) package, enabling an impressive computational throughput of up to $1\times10^{7}$ 
atom-steps per second on a single GPU.\cite{xu_gpumd_2025} 
Considering its overall performance, NEP stands out as one of the most promising 
MLIPs for large-scale simulations of \ga\ under a wide range of physical conditions.

The radial function $g_n(r_{ij})$, which serves as the basic building block of the NEP
descriptor, is defined as
\begin{equation}
g_n(r_{ij}) = \sum_{k} c^{ij}_{nk} f_k(r_{ij}),
\end{equation}
where $r_{ij}$ is the distance between atoms $i$ and $j$, 
$c^{ij}_{nk}$ is a trainable parameter, and $f_k(r_{ij})$ is a Chebyshev-based
radial basis function with a finite cutoff:
\begin{equation}
f_k(r_{ij}) = \frac{1}{2}\!\left[T_k\!\!\left(2(r_{ij}/r_{\mathrm{cut}} - 1)^2 - 1\right)+1\right] s_c(r_{ij}),
\end{equation}
where $T_k(x)$ is the $k$-th-order Chebyshev polynomial of the first kind, and 
$s_c(r_{ij})$ is a cutoff function
\begin{equation}
s_c(r_{ij}) =
\begin{cases}
  \frac{1}{2}\!\left[1+\cos(\pi r_{ij}/r_{\mathrm{cut}})\right], & r_{ij} \le r_{\mathrm{cut}}, \\
  0, & r_{ij} > r_{\mathrm{cut}}.
\end{cases}
\end{equation}

The $n$-th radial descriptor of atom $i$ is then constructed as
\begin{equation}
q_n^i = \sum_{j \ne i} g_n(r_{ij}),
\end{equation}
and the $n$-th angular descriptor of atom $i$ with angular-momentum index $l$
is defined as
\begin{equation}
q_{n,l}^i = \sum_{j \ne i}\sum_{k \ne i} g_n(r_{ij})~g_n(r_{ik})~P_l(\theta_{ijk}),
\end{equation}
where $P_l(\theta_{ijk})$ is the $l$-th-order Legendre polynomial of the angle
$\theta_{ijk}$ between atoms $i$, $j$, and $k$.

All descriptors are concatenated into a single vector, which is then passed
through a feedforward neural network to predict the atomic potential energy:
\begin{equation}
U^i =
\sum_{\mu=1}^{N_{\mathrm{neu}}}
\omega^{(1)}_\mu
\tanh\!\left(
\sum_{\nu=1}^{N_{\mathrm{des}}}
\omega^{(0)}_{\mu\nu} q^i_\nu - b^{(0)}_\mu
\right)
- b^{(1)},
\end{equation}
where $N_{\mathrm{neu}}$ is the number of neurons in the hidden layer,
$N_{\mathrm{des}}$ is the number of descriptors,
$\omega^{(1)}_\mu$ and $b^{(1)}$ are the output-layer weight and bias,
$\omega^{(0)}_{\mu\nu}$ and $b^{(0)}_\mu$ are the hidden-layer weights and biases,
and $q^i_\nu$ denotes the $\nu$-th descriptor of atom $i$.

\subsection{Hyperparameters}

In this work, we use the fourth generation of NEP\cite{song_general-purpose_2024} to construct 
the MLIP for \ga. To faithfully capture the complex potential energy landscape of \ga, 
we adopt a relatively large parameterization strategy. 
Specifically, 9 radial descriptors ($r_\text{cut}=6~\text{\AA}$) and 
42 angular descriptors ($r_\text{cut}=4~\text{\AA}$) are employed, 
with a single hidden layer containing 100 neurons. 
The maximum Chebyshev polynomial orders are set to 12 and 10 for the radial and angular 
descriptors, respectively. 
In addition, 4-body and 5-body angular terms are also included.\cite{fan_gpumd_2022} 
In total, the resulting model contains 11,377 trainable parameters. 
We also tested larger parameter settings; however, they did not provide noticeable improvements 
in accuracy while causing a substantial increase in computational cost.

A robust training-weighting scheme is needed to balance the accuracy across distinct atomic 
environments, which is essential for achieving successful training over an extremely wide range 
of average atomic energies, spanning from the most stable $\beta$ phase to highly unstable, nonstoichiometric 
amorphous states under high pressure. 
Two distinct training strategies are tested in this work. 
The first strategy is to adopt the phase-dependent scaling scheme proposed in Ref.~\onlinecite{zhao_complex_2023}, 
which enhances the descriptive capability of the model for key \ga\ phases, including its 
metastable crystalline structures. 
The second strategy is to scale the training weights of all configurations according to their 
average potential energy per atom, ensuring that the NEP model primarily focuses on low-energy 
configurations while still retaining reasonable descriptive capability for high-energy states.
The scaling factor $\mathcal{S}$ is defined as
\begin{equation}
\label{eq:scaling}
\mathcal{S} = \frac{1}{\left|e_i/\epsilon - e_{\beta\text{-phase}}/\epsilon\right|^{\alpha} + 1},
\end{equation}
where $e_i$ is the average potential energy per atom of the $i$-th configuration, 
$e_{\beta\text{-phase}}$ denotes that of the $\beta$ phase, 
and $\alpha$ and $\epsilon$ are small constants controlling the strength of the scaling.

The average energy, atomic forces, and virial of each training configuration are used as training targets, 
with global weights of $\lambda_e=\lambda_f=\lambda_v=1$. 
Both $\mathcal{L}_1$ and $\mathcal{L}_2$ regularizations are applied with $\lambda_1=\lambda_2=0.05$ to 
avoid overfitting.\cite{fan_gpumd_2022} 
In practice, we also shift the energy baseline of the training configurations to align with the 
isolated-atom energy predicted by NEP89\cite{liang_nep89_2025}, as we find that NEP performs much 
better under this shifted baseline. 

\subsection{Datasets}

The dataset reported in Ref.~\onlinecite{zhao_complex_2023}, which was originally used to construct 
soapGAP and tabGAP models for \ga\ and is hereafter referred to as the GAP dataset, 
serves as the main part of our training set.
The GAP dataset contains 1,630 configurations covering a wide range of Ga--O systems, 
from crystalline \ga\ to nonstoichiometric, amorphous, and molten states, 
and also includes a number of few-atom systems for calibration.
This dataset provides a rich collection of representative atomic configurations, although with 
a relatively limited number of samples. 
It is generally sufficient for broad, general-purpose applications where ultimate accuracy 
is not the primary concern.

However, the GAP dataset lacks certain key configurations required 
to recover essential physical features in our irradiation simulations. 
To address this issue, we augment the dataset with additional training configurations generated 
using exactly the same sampling protocol as for the GAP dataset.
These newly sampled configurations are found to lie beyond the extrapolation capability of tabGAP 
and are crucial for accurately capturing SHI irradiation-induced phenomena.
Two types of configurations are added. 
The first type comprises 60 structures along the energy--volume 
equation-of-state curve of the $\gamma$ phase. 
This addition is motivated by the fact that 
the $\beta \to \gamma$ transition is widely observed in radiation experiments on \ga,\cite{zhao_crystallization_2025,han_electronic_2025,abdullaev_ions_2025} 
yet $\gamma$-phase configurations are underrepresented in the original GAP dataset. 
The second type consists of 300 configurations sampled from heating--cooling processes 
of the $\beta$ phase under various pressure and volume conditions,
because such thermodynamic pathways frequently occur during irradiation.
Our principle component analysis (PCA) shows that these newly added configurations 
effectively fill the gap in the configuration space with the $\beta$ phase 
and the molten/amorphous states that exists in the original GAP dataset [Fig.~\ref{fig_melting_pca}(d)].

\subsection{Simulations}

The SHI irradiation simulations are performed using the GPUMD package,\cite{xu_gpumd_2025} 
following the strategy proposed in Ref.~\onlinecite{rymzhanov_damage_2017}.
In general, the simulations proceed as follows.
(1) The energy deposition from SHIs in \ga\ is simulated using TREKIS, a Monte Carlo code
developed to model the electronic kinetics following SHI impact on matter.\cite{medvedev_time-resolved_2015,rymzhanov_effects_2016} 
(2) The resulting radial energy profile of the ion track is used to initialize the atomic
velocities of the \ga\ lattice for subsequent MD simulations, with a conversion factor of 
0.45 applied to reproduce the experimentally observed track dimensions.
(3) The MD simulations are carried out in the NVE ensemble, while the boundary temperature
is controlled using Nose-Hoover chain thermostats, with the coupling parameter tuned to 1,000. 
Further methodological details regarding TREKIS are provided in Appendix~\ref{app:trekis}, 
and technical details for MD simulations are provided in Appendix~\ref{app:md}.

To ensure a fair comparison, all prediction tests for both NEP and tabGAP are performed 
using the same LAMMPS installation and hardware, eliminating potential software- and 
hardware-related biases.

\begin{figure}[htbp]
\centering
\includegraphics[width=0.7\columnwidth]{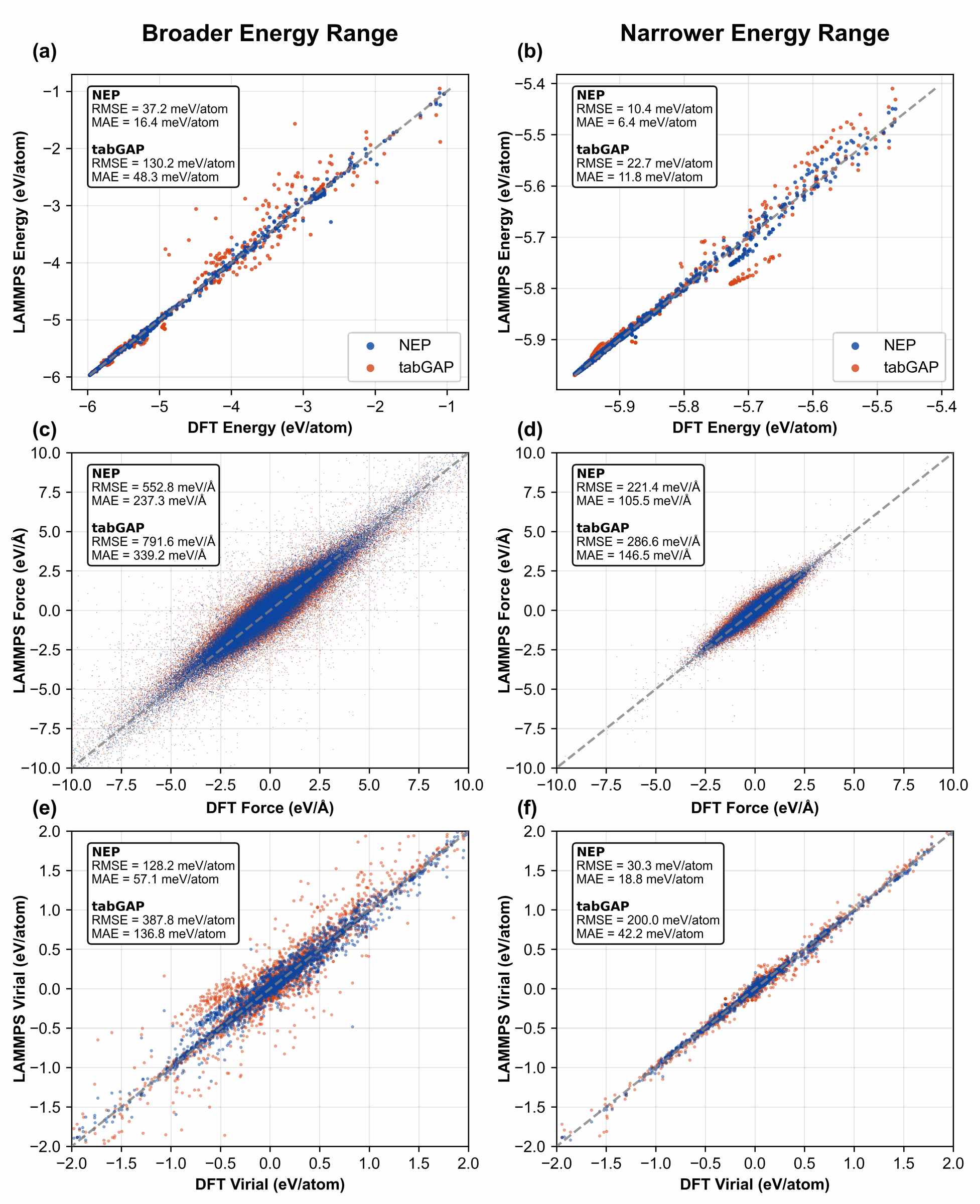}
\caption{\label{fig_comparison_dual_2x3}
Detailed comparison of the prediction accuracy of the NEP and tabGAP models.
Blue and red dots represent NEP and tabGAP results, respectively.
Panels (a), (c), and (e) show the predicted energies, forces,
and virials of the GAP-dataset configurations with average energies ranging from
$-5.9$ to $-0.9$~eV/atom.
Panels (b), (d), and (f) present the corresponding results
for a narrower energy window from $-5.9$ to $-5.4$~eV/atom.
}
\end{figure}

\section{RESULTS AND DISCUSSION}

\subsection{Overall performance comparison}

A comprehensive comparison of the predictive capabilities of NEP and tabGAP is
shown in Fig.~\ref{fig_comparison_dual_2x3} with results summarized in Table~\ref{tab_accuracy_comparison}. 
Our NEP model, trained with the
energy-dependent weighting strategy, exhibits consistently superior performance
over tabGAP.
Given the wide energy span of the original GAP dataset, we separately analyze
(1) configurations spanning a broad energy range up to 5~eV/atom above the $\beta$ phase 
(91.5\% of the dataset) and (2) relatively stable configurations within 
0.5~eV/atom above the $\beta$ phase (45.3\% of the dataset).

For the configurations with a broad energy range, NEP offers pronounced accuracy
improvements in predicting energies
[Fig.~\ref{fig_comparison_dual_2x3}(a)],
forces
[Fig.~\ref{fig_comparison_dual_2x3}(c)],
and virials
[Fig.~\ref{fig_comparison_dual_2x3}(e)]
compared with tabGAP.
Specifically, the mean absolute error (MAE) of the energy predicted by NEP is
16.4~meV/atom, about one third of that of tabGAP (48.3~meV/atom).
Moreover, the root mean square error (RMSE) of energy predicted by tabGAP
reaches 130.2~meV/atom, which is 3.5 times higher than that of NEP
(37.2~meV/atom), indicating a potential instability of tabGAP over the wide
energy range.
The MAE of force prediction by NEP is 237.3~\meVA, approximately 70\% 
of that of tabGAP (339.2~\meVA), while the MAE of virial prediction is
57.1~meV/atom for NEP and 136.8~meV/atom for tabGAP.
The overall MAEs of both models remain relatively large mainly because the dataset
covers an extremely broad energy range.
For the relatively stable configurations, both models show significant improvements in
energy 
[Fig.~\ref{fig_comparison_dual_2x3}(b)],
force
[Fig.~\ref{fig_comparison_dual_2x3}(d)]
and virial
[Fig.~\ref{fig_comparison_dual_2x3}(f)]
predictions.
The MAE of energy prediction by NEP is 6.4~meV/atom, about half of that of
tabGAP (11.8~meV/atom), and NEP still maintains noticeable advantages in force
and virial predictions.
Additional prediction accuracy comparisons over the entire GAP dataset and 
25.6\% of the GAP dataset (up to 0.1~eV/atom above the $\beta$ phase) are provided in Fig.~S1, 
confirming that the superiority of NEP in prediction accuracy is maintained consistently.
We also compare isolated Ga--Ga, Ga--O, and O--O dimer energy curves on both linear and logarithmic
energy scales, showing that NEP and tabGAP exhibit broadly comparable behavior in the pair-interaction
and highly repulsive short-range regimes (Figs.~S2 and S3).

\begin{table}[htbp]
\caption{\label{tab_accuracy_comparison}
Prediction accuracy comparison between tabGAP and NEP for configurations with 
broader (up to 5~eV/atom above $\beta$ phase) and narrower (up to 0.5~eV/atom 
above $\beta$ phase) energy ranges. Energy values are in meV/atom, force values 
are in meV/\AA, and virial values are in meV/atom.
}
\begin{ruledtabular}
\begin{tabular}{lccccccc}
 & \multicolumn{2}{c}{Energy} & \multicolumn{2}{c}{Force} & \multicolumn{2}{c}{Virial} \\
 \cline{2-3} \cline{4-5} \cline{6-7}
Model & MAE & RMSE & MAE & RMSE & MAE & RMSE \\
\hline
tabGAP (Broader) & 48.3 & 130.2 & 339.2 & 791.6 & 136.8 & 387.8 \\
NEP (Broader) & 16.4 & 37.2 & 237.3 & 552.8 & 57.1 & 128.2 \\
tabGAP (Narrower) & 11.8 & 22.7 & 146.5 & 286.6 & 42.2 & 200.0 \\
NEP (Narrower) & 6.4 & 10.4 & 105.5 & 221.4 & 18.8 & 30.3 \\
\end{tabular}
\end{ruledtabular}
\end{table}

\begin{figure}[htbp]
\centering
\includegraphics[width=0.7\columnwidth]{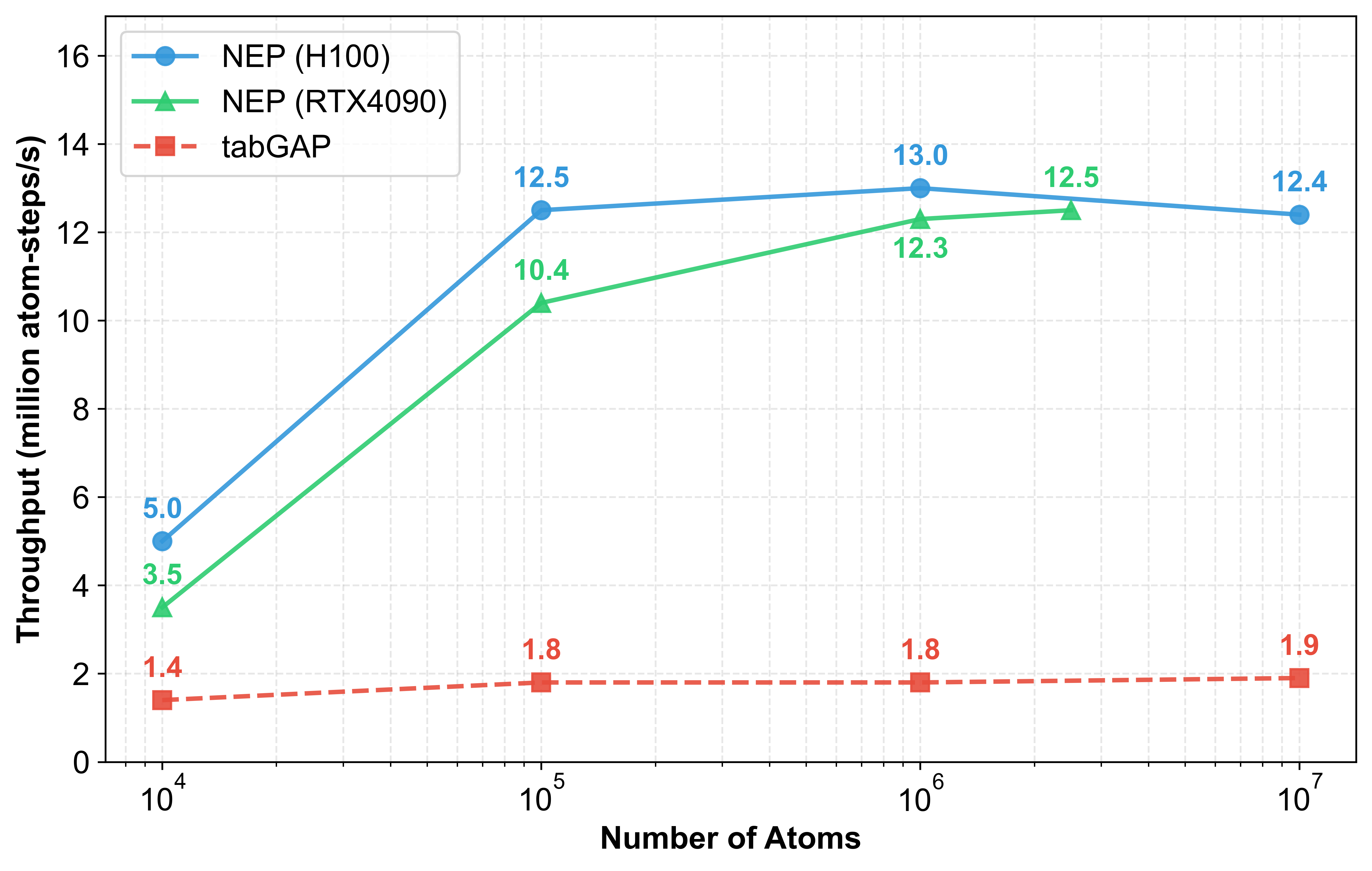}
\caption{\label{fig_throughput_comparison}
Computational throughput benchmark for NEP and tabGAP across different system sizes.
NEP is tested on a single NVIDIA H100 GPU and a single NVIDIA GeForce RTX~4090 GPU,
while tabGAP is tested on eight CPU nodes, each equipped with two 24-core Intel Xeon
Gold 6240R processors.
}
\end{figure}

Beyond its prediction accuracy, NEP also exhibits remarkably high computational throughput.
To characterize its performance, we benchmarked the throughput of NEP on two representative
GPU platforms---a single NVIDIA H100 (a current-generation data-center GPU) and a single
NVIDIA GeForce RTX~4090 (a high-end consumer-grade GPU)---alongside tabGAP running in LAMMPS
on a cluster of eight CPU nodes, each equipped with two 24-core Intel Xeon Gold 6240R
processors (released in 2020), for systems containing between $10^4$ and $10^7$ atoms
[Fig.~\ref{fig_throughput_comparison}].
On the H100, NEP reaches approximately $5.0\times10^6$ atom-steps per second at $10^4$ atoms
and approaches a stable throughput of around $1.25\times10^7$ atom-steps per second for
systems larger than $10^5$ atoms.
The RTX~4090 delivers slightly lower but comparable performance, reaching approximately
$1.04\times10^7$ atom-steps per second at $10^5$ atoms and $1.25\times10^7$ at $2.5\times10^6$ atoms,
although its 24~GB memory becomes a limiting factor for systems on the order of $10^7$ atoms.
In comparison, the tabGAP throughput on the eight-node CPU cluster plateaus at approximately
$2\times10^6$ atom-steps per second for large systems.
We acknowledge that newer-generation CPUs would yield higher throughput, and that LAMMPS
offers a degree of GPU acceleration for certain potential styles.
Nonetheless, the fact that even a single consumer-grade GPU can sustain a throughput
exceeding $10^7$ atom-steps per second highlights the substantial cost-effectiveness of
NEP natively deployed within GPUMD---a molecular dynamics package written entirely in CUDA
and designed for GPU execution from the ground up.

\begin{figure*}[htbp]
\centering
\includegraphics[width=\textwidth]{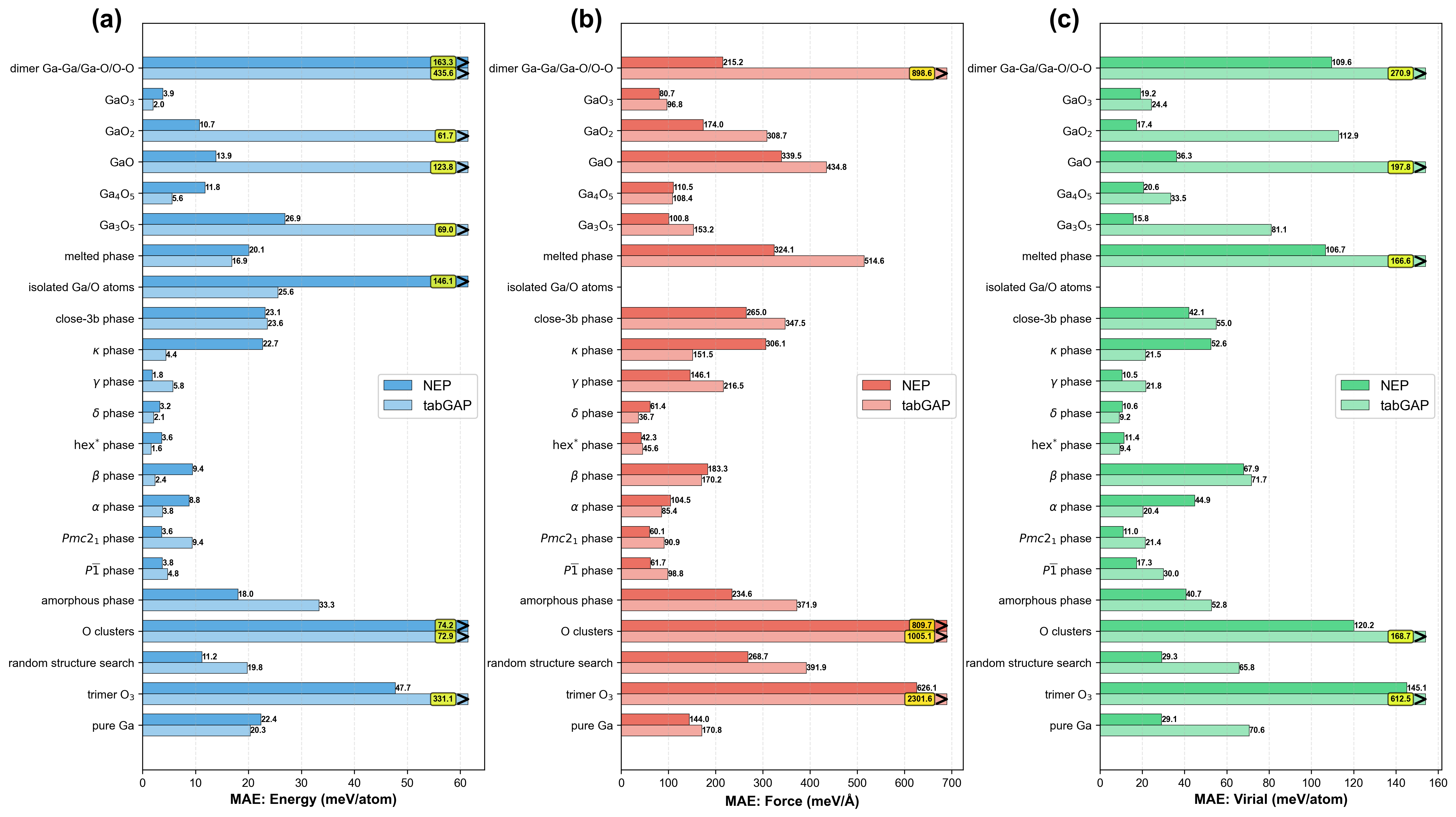}
\caption{\label{fig_mae_by_config_type_comparison}
Comparison of the MAE of (a) energy, (b) force, and (c) virial 
predictions by NEP and tabGAP for different config types in the GAP dataset.
}
\end{figure*}

\begin{figure}[htbp]
\centering
\includegraphics[width=0.7\columnwidth]{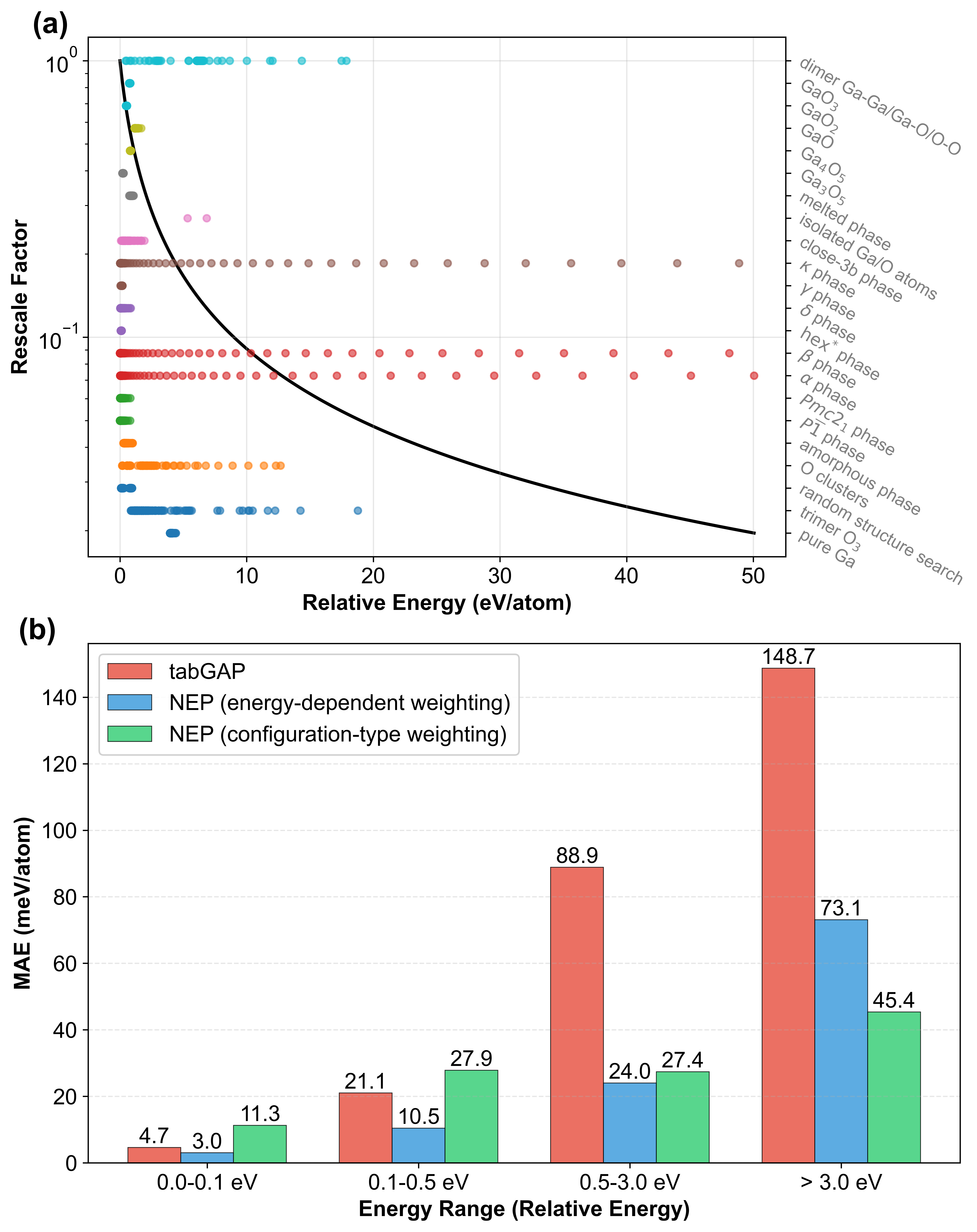}
\caption{\label{fig_energy_mae_comparison}
(a) Illustration of the energy-dependent weighting strategy, where the black curve
represents the rescaling factor as a function of the average potential energy.
All samples in the GAP dataset are shown as dots colored according to their configuration
types; their positions along the horizontal axis indicate their average potential energies.
(b) Comparison of the energy-prediction performance of tabGAP, NEP with
energy-dependent weighting, and NEP with configuration-type weighting over different
energy ranges.
}
\end{figure}

To elucidate the origins of this performance advantage, we systematically compared 
the prediction errors of NEP and tabGAP across all 22 configuration types in the GAP dataset
[Fig.~\ref{fig_mae_by_config_type_comparison}].
We find that the energy MAE of tabGAP exhibits strong fluctuations among different configuration
types, with some types showing very small errors below 3~meV/atom (e.g., $\beta$ phase,
$\delta$ phase, hex$^{*}$ phase, and $\text{GaO}_3$), while others display much larger errors
exceeding 50~meV/atom (e.g., dimers, trimers, O clusters, and various nonstoichiometric
structures).
This fluctuation is a direct consequence of the configuration-type weighting strategy adopted
in tabGAP, which assigns higher weights to crystalline phases of \ga.\cite{zhao_complex_2023} 
However, this strategy assigns identical weights to configurations that possess very different
average energies, as illustrated in Fig.~\ref{fig_energy_mae_comparison}(a).
For example, a $\beta$-phase crystal cell equilibrated at ambient conditions is assigned the
same weight as a $\beta$-phase crystal cell compressed to about 60\% of its lattice constant,
even though their average energies differ by $\sim$50~eV/atom.
From a thermodynamic perspective, configurations with higher average energies correspond to
significantly lower occurrence probabilities, whereas the configuration-type weighting strategy 
treats them equally.

In contrast, the NEP model demonstrates more consistent energy prediction accuracy
across different configuration types.
As shown in Fig.~\ref{fig_mae_by_config_type_comparison}, the NEP model trained with our 
energy-dependent weighting strategy 
substantially reduces the MAE for nonstoichiometric and amorphous
configurations while maintaining low MAE values for crystalline phases.
As a result, NEP achieves a markedly improved overall energy-prediction accuracy compared with
tabGAP across different energy ranges
[Fig.~\ref{fig_energy_mae_comparison}(b)].
Moreover, the improvement in force predictions achieved by NEP over tabGAP is less dramatic
but more uniform across different configuration types
[Fig.~\ref{fig_mae_by_config_type_comparison}(b)].
For noncrystalline configurations, the force MAE of NEP is consistently slightly smaller
than that of tabGAP, whereas for crystalline configurations the performances of the two
models are largely comparable.
Notably, the improvement in virial prediction is the most pronounced
[Fig.~\ref{fig_mae_by_config_type_comparison}(c)].
For noncrystalline configurations, the virial MAE of NEP is significantly smaller than that
of tabGAP, and for crystalline configurations NEP also consistently yields smaller errors.

We believe that the overall superior performance of NEP over tabGAP in predicting energies,
forces, and virials is primarily attributed to our energy-dependent weighting strategy,
as illustrated in Fig.~\ref{fig_energy_mae_comparison}(a).
In practice, the weighting factors are chosen as $\alpha = 1.5$ and $\epsilon = 1$~eV/atom.
The weighting factors remain close to unity for the most stable configurations, decrease
to 1/2 when the average potential energy increases by 1~eV/atom above that of the
optimized $\beta$ phase, and then rapidly drop with increasing energy, eventually becoming
smaller than 0.02 for the highest-energy configurations in the GAP dataset.
The energy-prediction performances of tabGAP, NEP trained with energy-dependent weighting,
and NEP trained with configuration-type weighting over different energy ranges are shown in
Fig.~\ref{fig_energy_mae_comparison}(b).
With configuration-type weighting, NEP outperforms tabGAP only for configurations whose
average energies exceed those of the optimized $\beta$ phase by more than 0.5~eV/atom.
In contrast, with the energy-dependent weighting scheme, NEP surpasses tabGAP across the
entire energy range, at the cost of only a decrease in accuracy for 
thermodynamically rare high-energy configurations.

\begin{figure}[htbp]
\centering
\includegraphics[width=0.7\columnwidth]{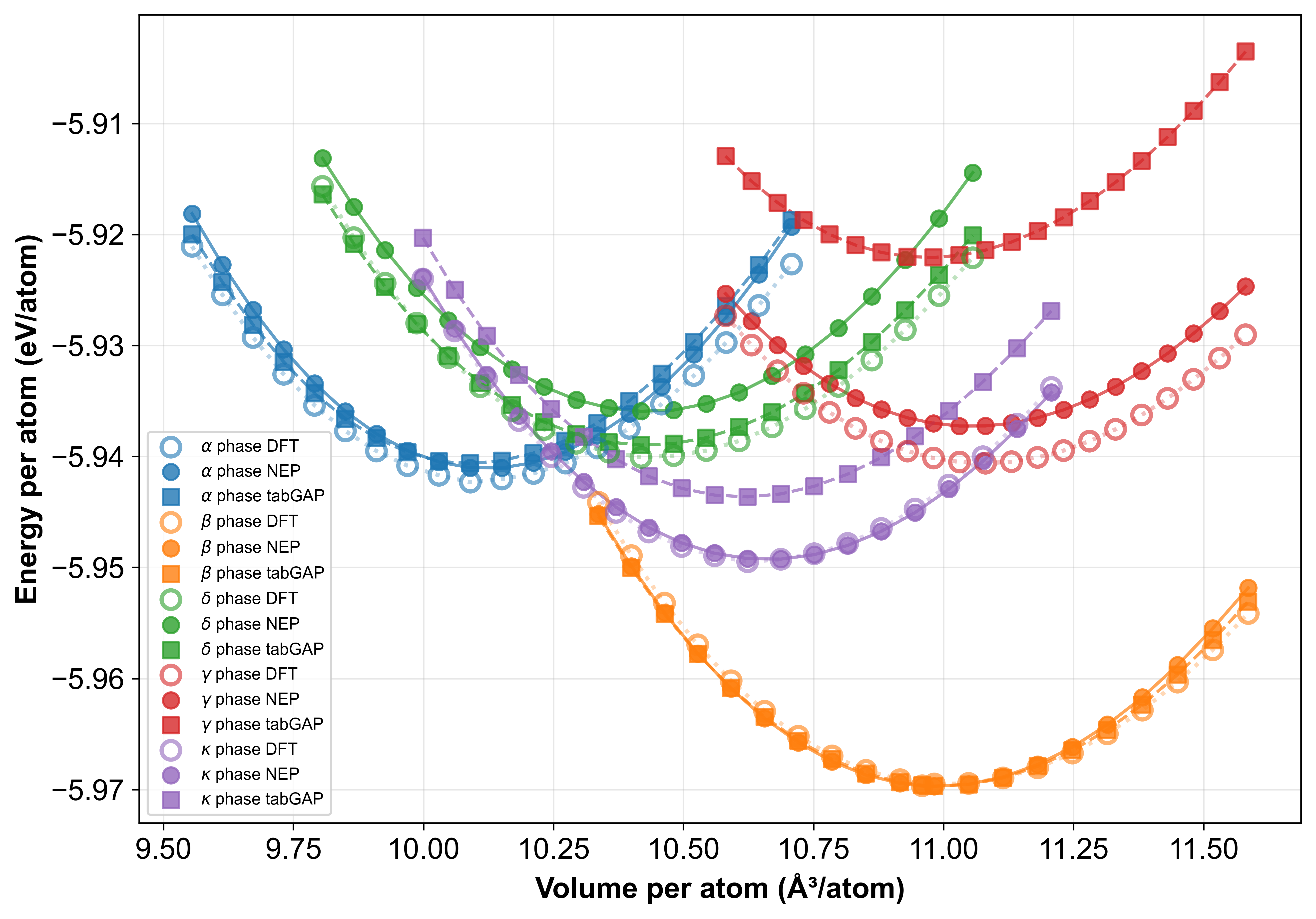}
\caption{\label{fig_energy_volume_comparison}
Energy-volume curves for the $\alpha$, $\beta$, $\delta$, $\gamma$, and $\kappa$ phases 
predicted by NEP (solid circles) and tabGAP (solid squares), compared with DFT reference 
data (open circles). Different phases are color-coded as indicated.
}
\end{figure}  

\begin{figure}[htbp]
\centering
\includegraphics[width=0.7\columnwidth]{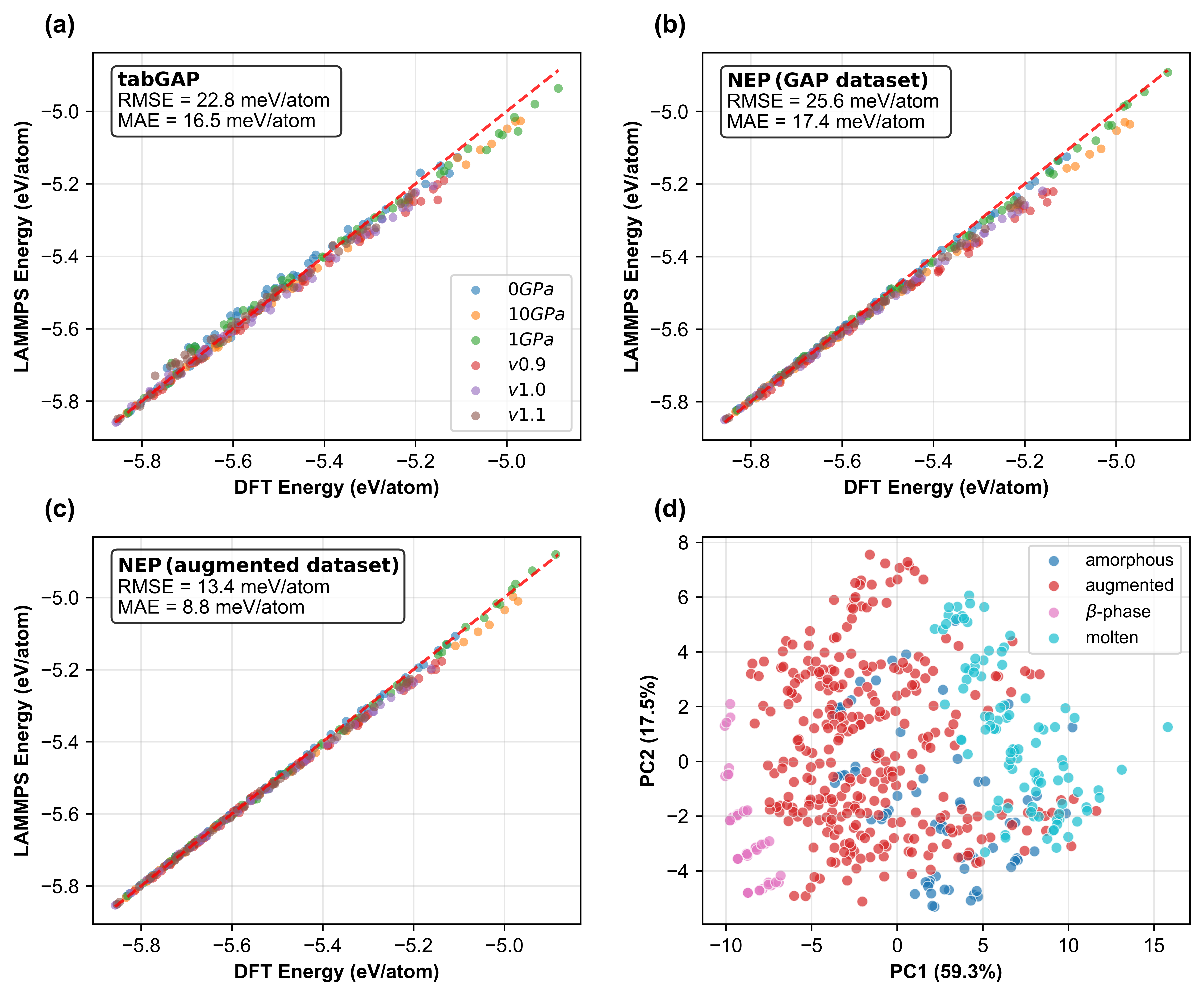}
\caption{\label{fig_melting_pca}
Energy predictions for configurations sampled from heating--cooling processes 
of the $\beta$ phase under various pressure and volume conditions, as predicted by (a) 
tabGAP, (b) NEP trained on the GAP dataset, and (c) NEP trained on the 
augmented dataset. 
(d) PCA projection of configurations from the heating--cooling processes 
along with related configurations of the GAP dataset.
}
\end{figure}

\subsection{Training set augmentation}

The $\beta \to \gamma$ transition is a key phenomenon widely observed in irradiation
experiments on \ga.\cite{zhao_crystallization_2025,han_electronic_2025,abdullaev_ions_2025} 
However, $\gamma$-phase configurations are underrepresented in the original GAP dataset,
raising concerns about whether tabGAP can accurately describe the physical properties of
the $\gamma$ phase.
To address this issue, we sampled 60 configurations along the energy--volume curve of the
$\gamma$ phase using the same sampling protocol as employed for the GAP dataset.
The energy--volume curve of the $\gamma$ phase obtained from DFT calculations, together
with the corresponding curves of the $\alpha$, $\beta$, $\delta$, and $\kappa$ phases
included in the original GAP dataset, are used as reference data in
Fig.~\ref{fig_energy_volume_comparison} to assess the performance of NEP (on the augmented dataset) 
and tabGAP in describing the equation of state of \ga\ polymorphs.

The first notable observation is that tabGAP systematically overpredicts the energy of
the $\gamma$ phase by nearly 20~meV/atom relative to the DFT results, indicating limited
extrapolation capability for zero-shot configurations that deviate from its training set.
NEP shows improved zero-shot performance for $\gamma$-phase configurations
[Fig.~S4], although it still overestimates their energies.
The NEP model trained with the augmented dataset performs best, achieving an energy MAE
about 3~meV/atom for the $\gamma$ phase.
In addition, both NEP and tabGAP accurately reproduce the energy--volume curves of the
$\alpha$ and $\beta$ phases, with predictions nearly overlapping with the DFT data.
These two phases are the most extensively studied polymorphs in the \ga\ community, and
their accurate equation-of-state description is therefore essential for the overall
reliability of MLIPs.
Notably, NEP also provides a good description of the $\kappa$-phase energy--volume curve,
whereas tabGAP exhibits a noticeable overestimation of the corresponding energies.
Given that the $\beta \to \kappa$ transition has also been reported in irradiation
experiments on \ga,\cite{han_electronic_2025} the improved description of the $\kappa$
phase constitutes another advantage of NEP over tabGAP.
It should be noted that the polymorph configurations used for the original GAP-dataset
energy--volume curves were not fully geometry-optimized.
To exclude possible effects associated with the lack of structural relaxation, we further optimized
these configurations using VASP and repeated the single-point DFT calculations and MLIP predictions.
The optimized-structure results lead to the same conclusions, confirming that the above comparison
is not affected by the initial lack of geometry optimization (Fig.~S5).

Although NEP performs better for most polymorphs, tabGAP shows a clear advantage in
describing the $\delta$ phase.
However, it is worth noting that both the $\delta$ and $\gamma$ phases are 
metastable structures that possess similar cohesive energies despite their different densities.
As a result, although NEP slightly overpredicts the energies of both $\delta$ and $\gamma$
phases by about 3~meV/atom, it still correctly captures their relative energetics and
thermodynamic similarity.
In contrast, the accurate description of the $\delta$ phase but the substantial
overprediction of the $\gamma$ phase by tabGAP suggest that tabGAP may fail to reproduce
the thermodynamic proximity between the $\delta$ and $\gamma$ phases.

Irradiation of solids is often accompanied by atomic recoil and/or electronic excitation
processes, which can give rise to pronounced local thermal effects.
As a consequence, melting, cooling, and subsequent recrystallization processes are
frequently observed during irradiation.
Whether an MLIP can reliably describe such highly non-equilibrium processes is therefore
crucial for irradiation simulations.
Moreover, the disordered structures generated during heating exhibit a much more complex
potential-energy landscape than the original crystalline phases, posing significant
challenges to both the completeness of the training dataset and the zero-shot capability
of MLIPs.

To systematically assess the performance of NEP and tabGAP in describing the
heating--cooling behavior of the $\beta$ phase, we sampled 600 configurations from six
independent MD simulations.
These simulations employed a NEP model trained on the original GAP dataset to perform 
heating--cooling cycles on a 160-atom $\beta$-phase supercell, with temperatures varying 
between 800~K and 4000~K. 
Six distinct simulation conditions were considered:

\noindent
(1) Config \textit{0GPa}: NPT at 0~GPa; \newline
(2) Config \textit{1GPa}: NPT at 1~GPa; \newline
(3) Config \textit{10GPa}: NPT at 10~GPa; \newline
(4) Config \textit{v1}: NVT at the original cell volume; \newline
(5) Config \textit{v0.9}: NVT with the volume rescaled to 0.9; \newline
(6) Config \textit{v1.1}: NVT with the volume rescaled to 1.1. 

The energy predictions obtained using tabGAP and the NEP model trained on the GAP dataset
are compared with the corresponding DFT reference data in
Fig.~\ref{fig_melting_pca}(a) and Fig.~\ref{fig_melting_pca}(b), respectively.
Overall, NEP exhibits better zero-shot performance for low-energy configurations than
tabGAP, whereas both models systematically underpredict the energies of high-energy
configurations.
Specifically, tabGAP overestimates the energies of the low-energy configurations in
Configs \textit{0GPa}, \textit{1GPa}, and \textit{v1.1}, indicating an overall energy
overestimation for configurations corresponding to relatively low environmental pressures
near the melting temperature.
In contrast, NEP consistently provides more accurate predictions for the low-energy
configurations, further demonstrating its superior extrapolation capability.
For the high-energy configurations, both models exhibit systematic softening 
issues,\cite{deng_systematic_2024} characterized by an overall underprediction
of the energies of configurations far from equilibrium.
This softening behavior is more pronounced for NEP than for tabGAP, resulting in an overall
MAE of 17.4~meV/atom in Fig.~\ref{fig_melting_pca}(b), slightly larger than that of tabGAP
(16.5~meV/atom) shown in Fig.~\ref{fig_melting_pca}(a).

To alleviate the softening problem of NEP, half of these 600 configurations were selected
based on PCA filtering\cite{chen_neptrain_2025} and added to the augmented training dataset.
As shown in Fig.~\ref{fig_melting_pca}(c), the NEP model trained on the augmented dataset
exhibits significantly improved performance across the entire energy range.
The overall MAE is reduced to 8.8~meV/atom, which is approximately half of that obtained with
models trained on the original GAP dataset.
Although the softening problem is not completely eliminated, this is a reasonable compromise 
by the energy-dependent weighting strategy, as the primary objective of our NEP training is 
to accurately describe stable and metastable states under ambient conditions.
Moreover, this dataset augmentation yields an unexpected benefit: 
an improved description of the anisotropic lattice thermal conductivity (LTC) of \ga. 
As shown in Fig.~S6, the LTC predicted by the augmented NEP potential correctly reproduces the 
pronounced preference for the [010] direction over the other two crystallographic directions, 
although the absolute LTC values remain systematically lower than experimental measurements, 
as the NEP was not specifically trained to target LTC.
Figure~\ref{fig_melting_pca}(d) presents the PCA projection of the training configurations
from the heating--cooling processes together with related configurations of the GAP
dataset.
The newly added samples effectively fill the gaps in descriptor space between the original
$\beta$-phase, molten, and amorphous configurations, thereby validating our sampling strategy.
In addition to the improved energy prediction shown in Fig.~\ref{fig_melting_pca}(c), we further
benchmark the NEP model trained on the augmented dataset using independent molten-state tests.
The noncrystalline energy--volume comparison includes both structures from the Zhao \textit{et al.}
training dataset and additional stretched or compressed molten-state configurations generated from
2200~K structures in this work.
The latter configurations, especially after geometry optimization, lie much closer to the
$\beta$-phase ground-state energy, where NEP follows the DFT reference data more closely than
tabGAP (Fig.~S7).
Structural benchmarks for 2200~K noncrystalline Ga$_2$O$_3$, including RDFs, bond-angle
distributions, and Ga--O ring-size statistics, further show that the augmented NEP model performs
comparably to or better than tabGAP across multiple molten-state descriptors (Figs.~S8--S10).

\begin{figure*}[htbp]
\centering
\includegraphics[width=\textwidth]{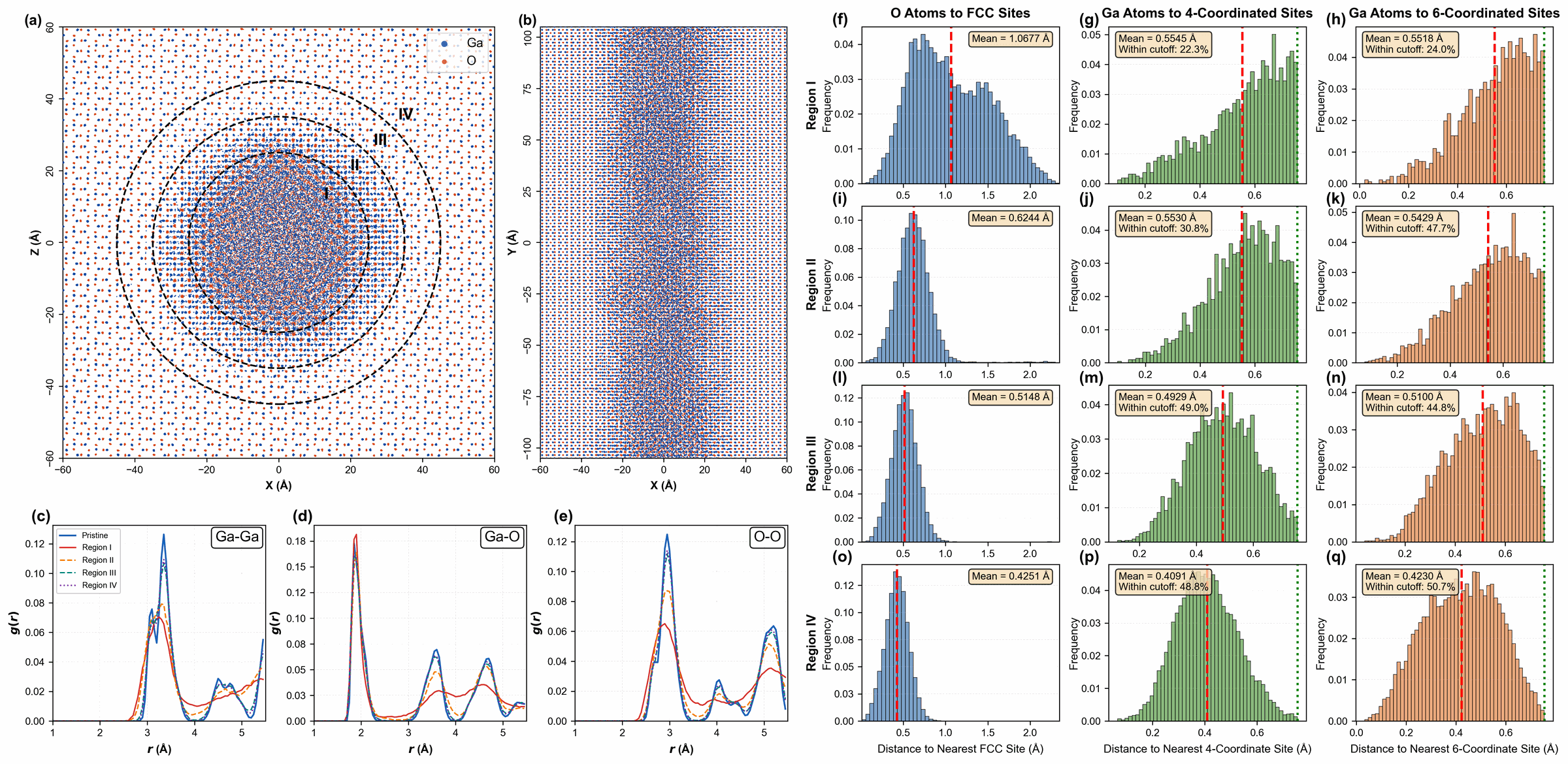}
\caption{\label{fig_ion_track}
Analysis of a representative ion track in $\beta$-\ga\ generated by the energy deposition from a 70~MeV 
\textsuperscript{181}Ta ion. 
(a) Transverse and (b) longitudinal cross-sectional views of the 
SHI-irradiated $\beta$-\ga\ atomic structure. 
Four distinct regions are identified in the transverse view (from center to periphery): 
amorphous core (region $\text{I}$), $\gamma$ phase (region $\text{II}$), 
deformed $\beta$ phase (region $\text{III}$), and pristine $\beta$ phase (region $\text{IV}$).
(c-e) Radial distribution functions for each region.
(f-q) Atomic-displacement distributions relative to the ideal $\beta$-phase 
lattice sites for each region.
}
\end{figure*}

\subsection{SHI irradiation simulations}

Recently, SHI irradiation experiments on $\beta$-\ga\ have been carried out by three
independent groups.\cite{xu_thermal_2025,han_electronic_2025,abdullaev_ions_2025,ai_radiation_2019} 
However, these studies have reported notably different conclusions regarding the
resulting track structures.
Xu \textit{et al.}\cite{xu_thermal_2025} reported that the tracks are fully 
amorphous when the electronic energy loss ($S_e$) exceeds a threshold of approximately 17~keV/nm.
In contrast, Han \textit{et al.}\cite{han_electronic_2025} observed the formation of
$\gamma$, $\delta$, and $\kappa$ phases within the tracks.
Abdullaev \textit{et al.}\cite{abdullaev_ions_2025} further reported that the tracks in
$\beta$-\ga\ are nearly pure $\gamma$ phase at $S_e$ of approximately 21~keV/nm.
To clarify these experimental discrepancies, reliable theoretical simulations of SHI
irradiation in $\beta$-\ga\ are highly desirable.
Although several related MD simulations based on tabGAP potentials have been reported,\cite{han_electronic_2025,abdullaev_ions_2025} 
the limitations of tabGAP discussed above raise
concerns regarding the reliability of its predictions for track structures.
Therefore, we performed SHI irradiation simulations on $\beta$-\ga\ using the NEP model
trained on the augmented dataset with energy-dependent weighting, with the aim of
demonstrating the practical predictive capability of NEP by elucidating the experimental discrepancies.

\begin{figure*}[htbp]
\centering
\includegraphics[width=\textwidth]{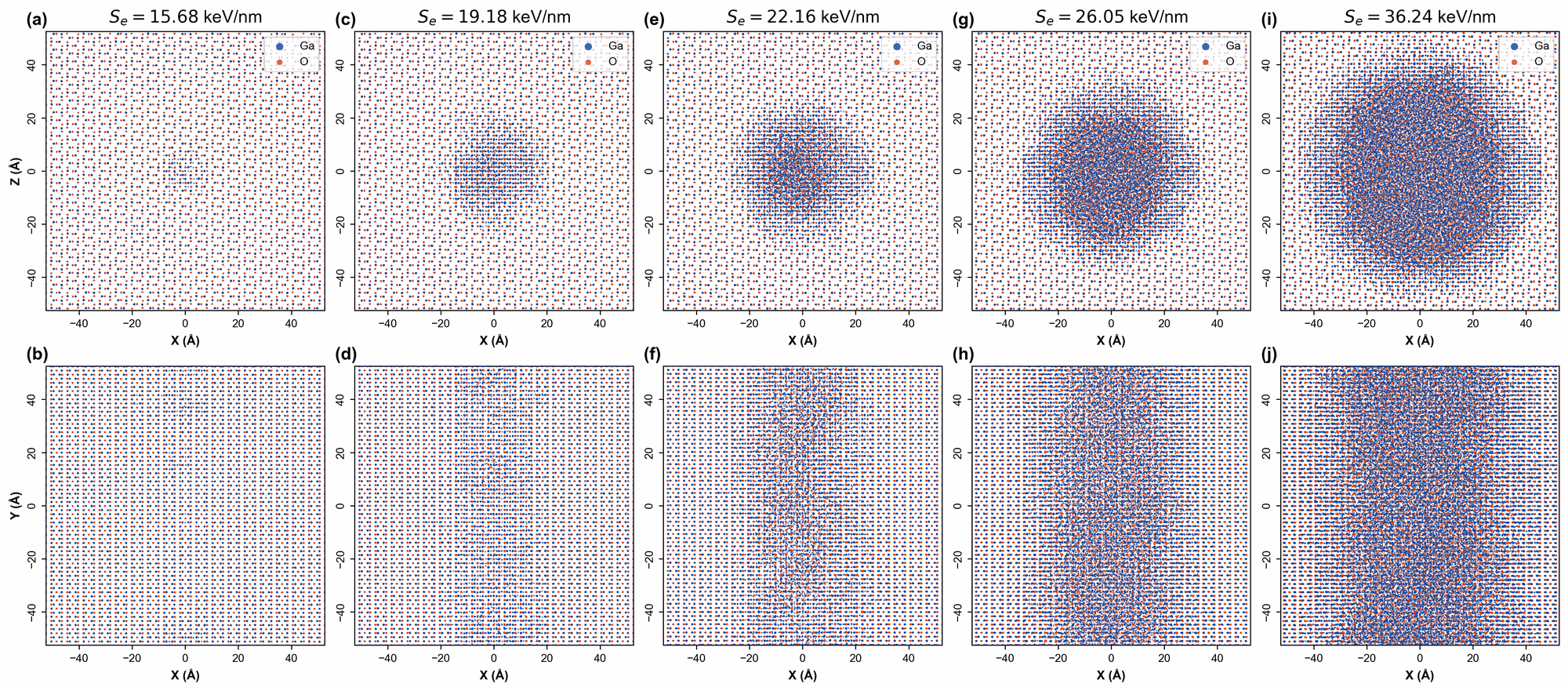}
\caption{\label{fig_tracks}
Transverse and longitudinal cross-sectional views of SHI-irradiated $\beta$-\ga\ 
at selected electronic energy losses:
(a--b) $15.68$~keV/nm, (c--d) $19.18$~keV/nm, 
(e--f) $22.16$~keV/nm, (g--h) $26.05$~keV/nm, and (i--j) $36.24$~keV/nm.
}
\end{figure*}

A representative ion track formed in $\beta$-\ga\ and the corresponding structural
analysis are shown in Fig.~\ref{fig_ion_track}.
The ion track is generated by MD simulations driven by the energy deposition of a
70~MeV $^{181}$Ta ion ($S_e = 26.05$~keV/nm in TREKIS), with a total simulation time of 200~ps.
Based on the transverse cross-sectional view
[Fig.~\ref{fig_ion_track}(a)], the atomic configuration is manually divided into four
concentric regions from the center to the periphery:
the amorphous core (region~\text{I}),
the $\gamma$ phase (region~\text{II}),
a deformed $\beta$ phase (region~\text{III}),
and the pristine $\beta$ phase (region~\text{IV}).
Figure~\ref{fig_ion_track}(c-e) presents the radial distribution functions (RDFs) for each
region, while Fig.~\ref{fig_ion_track}(f-q) shows the distributions of atomic displacements
with respect to the ideal $\beta$-phase lattice sites; together, these analyses are used
to assist structural identification. 
More conventional crystalline-structure analysis methods are not adopted here because
the atomic structure of the $\gamma$ phase remains highly disordered and is not uniquely
defined.\cite{ratcliff_tackling_2022} 
Although the $\gamma$-\ga\ phase is commonly associated with a defect-spinel structure,
it can be more naturally interpreted as a disordered form of $\beta$-\ga.\cite{huang_nature_2024} 
Importantly, during the $\beta \to \gamma$ transition, the oxygen sublattice remains
largely preserved. 
This allows the oxygen lattice of $\beta$-\ga\ to be used as a common reference framework
for structural analysis.
From this perspective, gallium atoms are expected to occupy the four- and six-coordinated
sites of the oxygen sublattice in both the $\beta$ and $\gamma$ phases, while the relative
population of these coordination environments differs between the two phases.
Therefore, the combined use of RDFs and atomic-displacement distributions provides a
robust and physically meaningful approach for identifying the structural characteristics
within the ion track.

According to the findings of Ref.~\onlinecite{abdullaev_ions_2025}, the $\gamma$ phase,
like the amorphous phase, also appears as a high-contrast region in transmission electron
microscopy (TEM) imaging. Including the $\gamma$-phase shell in the track size estimation,
the ion track shown in Fig.~\ref{fig_ion_track} has a diameter of approximately 5.9~nm,
consistent with the TEM measurements reported in Ref.~\onlinecite{xu_thermal_2025}.
In the core region, the RDFs are characterized by a solitary nearest-neighbor peak, 
with the absence of long-range structural features confirming the transition to a disordered state.
The atomic-displacement distributions further reveal that oxygen atoms completely leave
their original fcc sites (Fig.~\ref{fig_ion_track}(f)), and most gallium atoms are also displaced away from the vicinity of
their original four- and six-coordinated sites (Fig.~\ref{fig_ion_track}(g-h)).
Outside the amorphous core, a clear $\gamma$-phase shell is observed.
In this region, gallium atoms are randomly distributed among four- and six-coordinated
sites, while oxygen atoms predominantly occupy their pristine fcc positions.
The RDF analysis shows the absence of the second gallium--gallium peak, whereas the characteristic
fcc oxygen peak gradually recovers.
As shown in Fig.~\ref{fig_ion_track}(i-k), most oxygen atoms in this region return to their
original fcc sites, and most gallium atoms also reoccupy four- and six-coordinated sites,
but with a coordination ratio of approximately 1:1.55, differing from the pristine
$\beta$-phase ratio of 1:1.
This ratio is more consistent with the experimentally reported value of 1:1.35 for the
$\gamma$-phase structure.\cite{playford_structures_2013} 

Beyond the $\gamma$-phase region, a deformed $\beta$-phase structure is identified.
Although the deformation is not readily apparent in the transverse view
[Fig.~\ref{fig_ion_track}(a)], it becomes evident from the RDFs and the atomic-displacement
distributions.
Specifically, Fig.~\ref{fig_ion_track}(l-n) shows an overall shift of both oxygen and gallium
displacement distributions in region~\text{III} relative to the pristine $\beta$-phase
lattice sites (region~\text{IV}), while the widths of these distributions remain comparable.
This indicates that region~\text{III} retains the $\beta$-phase structure but experiences 
deformation induced by the ion track.
Such deformation has also been observed in our TEM characterizations and is found to
significantly influence the electronic properties of the $\beta$ phase; these results will
be reported elsewhere.

\begin{figure}[htbp]
\centering
\includegraphics[width=0.7\columnwidth]{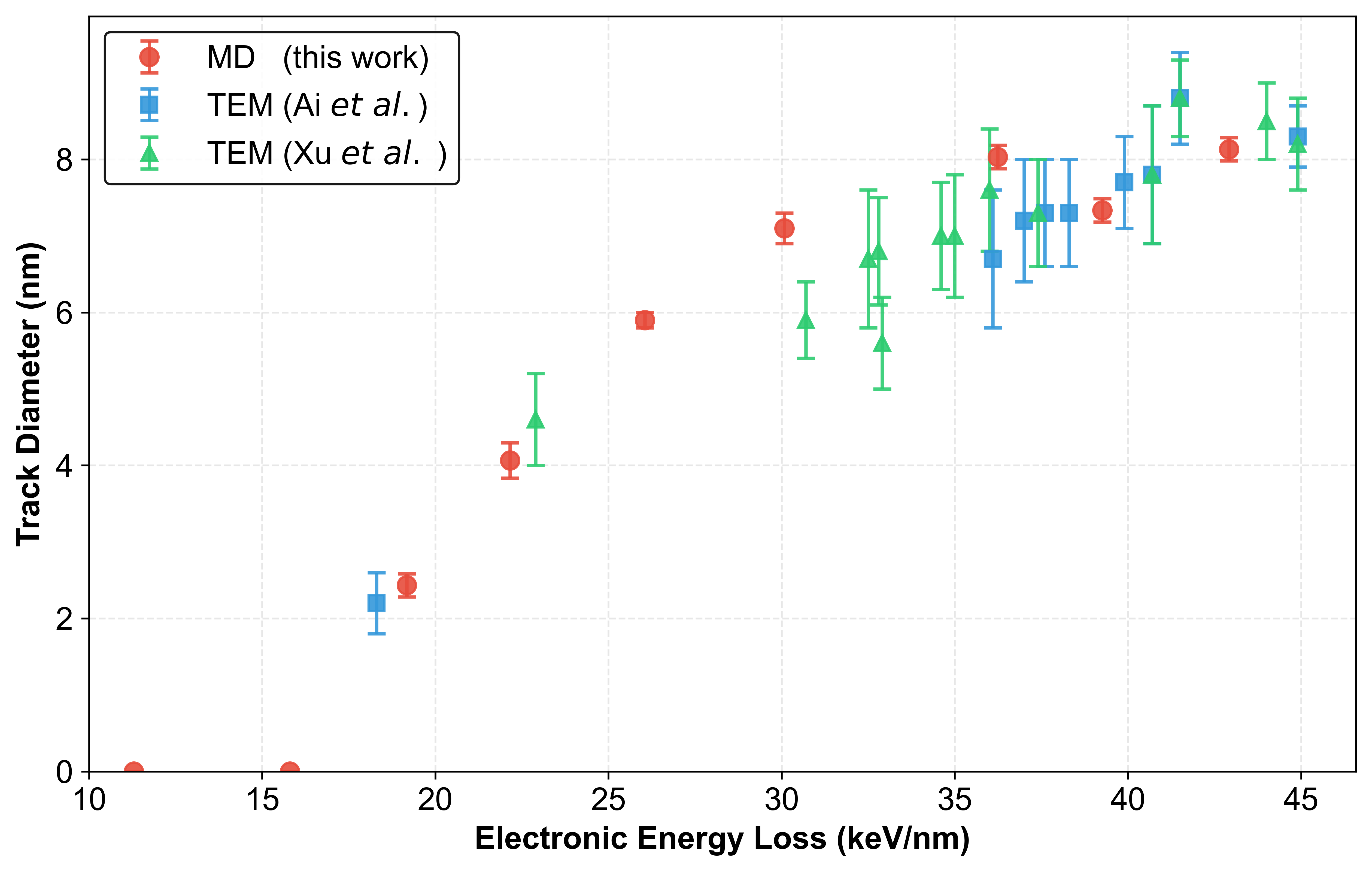}
\caption{\label{fig_track_diameter_comparison}
Comparison of the track diameters predicted in this work and the TEM 
results reported by Xu \textit{et al.}~\cite{xu_thermal_2025} 
and Ai \textit{et al.}~\cite{ai_radiation_2019}.}
\end{figure}

One of the central controversies in recent SHI irradiation experiments on
$\beta$-\ga\ concerns the crystallinity of the ion tracks.
Xu \textit{et al.}~\cite{xu_thermal_2025} reported that the tracks are fully amorphous,
whereas Abdullaev \textit{et al.}~\cite{abdullaev_ions_2025} concluded that the tracks
consist predominantly of the $\gamma$ phase.
It is noteworthy that the samples studied by Abdullaev \textit{et al.} were irradiated
with a relatively low $S_e$ of approximately 21~keV/nm, whereas most
samples examined by Xu \textit{et al.} were subjected to higher electronic energy losses
exceeding 30~keV/nm.
This difference in $S_e$ may therefore provide a natural explanation
for the seemingly conflicting experimental conclusions.
However, Xu \textit{et al.} also reported amorphous tracks at energy losses close to
20~keV/nm, which appears difficult to reconcile directly with the results of
Abdullaev \textit{et al.}
This situation motivates a systematic investigation of SHI irradiation over a wide
range of $S_e$.
To this end, we performed a series of SHI irradiation simulations with $S_e$ ranging 
from 11.29 to 42.92~keV/nm and 
five representative cases are shown in Fig.~\ref{fig_tracks}.

When $S_e = 15.68$~keV/nm, no ion track is formed, consistent with the experimental
threshold of approximately 17~keV/nm reported by Xu \textit{et al.}~\cite{xu_thermal_2025}.
The oxygen sublattice remains largely intact and defect-free, in agreement with
previous theoretical predictions.\cite{he_ultrahigh_2024} 
Only a small number of gallium interstitials are generated, most of which occupy
four- or six-coordinated sites within the FCC oxygen framework.
Such defect configurations can be regarded as incipient signatures of $\gamma$-phase
formation.
When $S_e$ increases to 19.18~keV/nm, slightly above the experimental threshold, 
continuous $\gamma$-phase regions are observed in the both transverse and longitudinal cross-sectional views, 
with limited number of oxygen defects emerging.
The simulated diffraction patterns for this case [Fig.~S11] reproduce the key features 
observed experimentally,\cite{abdullaev_ions_2025} confirming the formation of $\gamma$-phase 
ion tracks at $S_e$ slightly above the threshold. 

At $S_e = 22.16$~keV/nm, corresponding to the energy-loss regime where the most
controversial experimental interpretations have been reported, a clear core--shell
track structure emerges.
The shell consists predominantly of a well-defined $\gamma$ phase, while the core
forms a highly disordered region.
Notably, although this core region is highly disordered, it is not fully amorphous.
A substantial fraction of oxygen atoms remain distributed around their pristine
FCC sites, albeit with large displacements.
This quasi-FCC oxygen framework partially constrains the displacement of gallium atoms.
Along the ion penetration direction, the thicknesses of the core and shell
fluctuate heavily, and in some segments the disordered core even disappears, resulting in
a track composed entirely of the $\gamma$ phase.
These observations suggest that this highly disordered core should be interpreted
as a precursor state of the $\gamma$ phase, which may transform into a standard 
$\gamma$ structure under mild post-irradiation conditions, such as
long-term defect relaxation at ambient conditions or TEM-induced structural recovery.
To illustrate this point, we performed an annealing simulation to accelerate the post-irradiation evolution 
of the track structure at $S_e = 22.16$~keV/nm, maintaining a temperature of 800~K for 2~ns. 
Radial distribution functions (RDFs) provide clear evidence for the transition of the core from an amorphous 
to a $\gamma$-phase state (Fig.~S12), and a movie of this process, produced by OVITO\cite{stukowskiVisualizationAnalysisAtomistic2010}, 
is available online (Multimedia available online).

When $S_e$ further increases to 26.05~keV/nm, the track size increases rapidly to 
approximately 6~nm in diameter.
In this regime, the track core becomes fully amorphous, exhibiting no recognizable
structural motifs.
Compared with the case at $S_e = 22.16$~keV/nm, fluctuations along the ion penetration
direction are strongly suppressed, indicating the formation of a continuous amorphous
core and a stable core--shell structure.
At $S_e = 36.24$~keV/nm, the track size increases further, with the track diameter
reaching approximately 8~nm, while the core--shell morphology remains evident.
Interestingly, the thickness of the surrounding $\gamma$-phase shell does not
increase with increasing track size and saturates at about 1~nm.

To validate our simulation results, we compared the predicted track diameters with
experimental measurements reported in Refs.~\onlinecite{xu_thermal_2025,ai_radiation_2019},
as shown in Fig.~\ref{fig_track_diameter_comparison}.
The figure includes simulation results for nine different values of $S_e$, with each
case repeated three times to reduce statistical fluctuations in the MD simulations. 
All simulated transverse cross-sectional views, along with the corresponding measured 
track diameters, are presented in Fig.~S13 and Fig.~S14. 
The simulated track diameter increases rapidly once $S_e$ exceeds a threshold of 
approximately 17~keV/nm, whereas the growth rate decreases markedly when $S_e$ exceeds 30~keV/nm.
At even higher $S_e$, the monotonic dependence of the simulated track size on 
$S_e$ becomes less evident, which we attribute to the velocity effect on the radial 
energy deposition profile.
As shown in Fig.~S15, although the $S_e$ of 1000~MeV Ta ions (39.26~keV/nm) is slightly 
higher than that of 200~MeV Ta ions (36.24~keV/nm), the higher velocity of the 
1000~MeV ions leads to a broader spatial distribution of the deposited energy, resulting 
in a lower energy density near the track center and consequently a comparable track size.
Similarly, 1000~MeV Au ions have a notably higher $S_e$ (42.92~keV/nm) than 200~MeV Ta 
ions, yet their radial energy deposition densities near the track center are nearly 
identical, which explains why the two cases yield similar track diameters.
In general, the excellent agreement between simulation and experiment across a wide range of $S_e$ 
demonstrates the reliability of our NEP model.

In addition to accuracy, the computational cost of such a systematic study is also worth
noting. The total computational workload for the ion-track simulations presented
in this work exceeds $1.2\times10^{13}$ atom-steps.
Based on our benchmarks, a four-GPU RTX~4090 workstation running four independent 
simulations concurrently can deliver an aggregate throughput of roughly $4\times10^7$ 
atom-steps per second, enabling this entire workload to be completed in approximately 
3.5~days on a consumer-grade platform.
This demonstrates that, with NEP natively deployed within GPUMD, large-scale and 
high-throughput MD studies become readily feasible even without dedicated 
high-performance computing resources.

\section{Conclusion}

In conclusion, by combining the NEP architecture with an energy-dependent weighting
strategy, we develop a robust, accurate, and computationally efficient MLIP for \ga,
which demonstrates clear and comprehensive advantages over the state-of-the-art tabGAP
potential.
Furthermore, by adopting a physically process-oriented sampling strategy, the NEP model
trained on the augmented dataset is successfully applied to SHI irradiation simulations
of $\beta$-\ga.
The simulated results are in quantitative agreement with experimental observations and
provide a consistent and reliable physical explanation for the reported experimental
discrepancies.
In addition, the impressive computational efficiency of NEP enables large-scale MD
simulations approaching the device scale for \ga\ systems.\cite{xu_device-scale_2025} 
This capability opens a new avenue for device-scale design and optimization of
\ga-based devices from an atomistic simulation perspective.

\section*{SUPPLEMENTARY MATERIAL}
The supplementary material includes fifteen figures (Figs.~S1--S15) providing additional 
details on the NEP training accuracy, dimer interaction tests, energy-volume curves,
geometry-optimization tests for polymorph energy-volume curves, lattice thermal conductivity,
molten-state benchmarks, ion track structure and diffraction, radial distribution functions,
transverse cross-sectional views of ion tracks, and radial energy deposition profiles.
Also included are the Ga$_2$O$_3$ complex dielectric function file (\texttt{Ga2O3.txt}) and 
input parameters file (\texttt{INPUT\_PARAMETERS.txt}) used for TREKIS calculations. 
A movie showing the amorphous-to-$\gamma$ transition of the track core during 
post-irradiation annealing at 800~K for 2~ns is also provided.

\begin{acknowledgments}
The authors acknowledge financial support from the National Natural Science Foundation of China
(Grant Nos.~12325511, 12405320, and 11905262),
the CAS Talent Program Youth Project of Haizou Xue,
the Huizhou Science and Technology Talent Project (Grant No.~2024EQ050015),
and the Gansu Provincial Science and Technology Innovation Talent Program (Grant No.~24RCKB011).
The authors also thank Jiaming Zhang and Wentao Wang for their valuable comments and suggestions.
\end{acknowledgments}

\section*{DATA AVAILABILITY}
All essential training, simulation and analysis scripts for this work are available on the 
GitHub repository: \url{https://github.com/amphilagus/Ga2O3-ion-track}.

\appendix

\section{Methodology of TREKIS}
\label{app:trekis}
The central parameter in TREKIS is the differential scattering cross-section 
$\sigma$, which describes the interactions between ballistic electrons and 
orbital electrons. 
This parameter is calculated using the DSF-CDF formalism, based 
on optical data obtained from photo-absorption experiments: 
\begin{equation} 
\frac{\partial^{2}\sigma}{\partial(\hbar\omega)\partial(\hbar q)} = 
\frac{2[Z_{\text{e}}(v,q)e]^{2}}{\pi\hbar^{2}v^{2}} \frac{1}{\hbar 
q}\mathrm{Im}[\frac{-1}{\epsilon(\omega,q)}]\ , \label{eq1} 
\end{equation} 
In this formalism, $\hbar\omega$ represents the transferred energy, $q$ denotes 
the transferred momentum, $v$ is the ion velocity, $Z_e$ is the effective 
charge of the ion passing through \ga, and $\epsilon$ is the complex 
dielectric function. 
The ion charge $Z_{\text{e}}(v,q)$ is calculated with Barkas formula:\cite{Gervais1994} 
\begin{equation} 
Z_{\text{e}}(v,q) = Z_{\text{ion}}(1-\exp(-\frac{v}{v_{0}}Z_{\text{ion}}^{-\frac{2}{3}})) 
\end{equation} 
where $v$ is the velocity of the ion, $v_{0}=c/125$ is the empirical atomic 
electron velocity and $Z_{\text{ion}}$ is the ion charge in full ionization. 
The optical energy loss function of \ga\ is taken from 
Ref.~\onlinecite{he_first-principles_2006}, and fitted with a finite sum of Drude-Lorentz 
oscillator functions: 
\begin{equation} 
\mathrm{Im}(\frac{-1}{\epsilon(\omega,0)}) = \sum_{i} 
\frac{A_{\mathrm{i}}\gamma_{i}\hbar\omega}{(\hbar^{2}\omega^{2}-E_{i}^{2})^{2}+(\gamma_{i}\hbar\omega)^{2}} 
\end{equation} 
where $A_{i}$, $\gamma_{i}$, and $E_{i}$ are the fitted parameters, which are 
constrained by two sum rules: 
\begin{enumerate} 
\item The f-sum rule: 
\begin{equation} 
\frac{2}{\pi 
\Omega_{\text{p}}^{2}}\int_{I_{\text{p}}}^{\infty}\mathrm{Im}(\frac{-1}{\epsilon(\omega,0)})_{\text{shell}}\omega 
\mathrm{d}\omega = N_{\text{e,shell}} 
\end{equation} 
where $\Omega_{\text{p}}^{2}=4\pi e^{2}n_{at}/m_{\text{e}}$ is the plasma frequency, 
$N_{\text{e,shell}}$ is the number of electrons in selected shell, and $I_{\text{p}}$ is 
ionization potential. 
\item The KK-sum rule: 
\begin{equation} 
\frac{2}{\pi}\int_{0}^{\infty} 
\mathrm{Im}(\frac{-1}{\epsilon(\omega,0)})\frac{d\omega}{\omega} = 1 
\end{equation} 
\end{enumerate} 
The free-electron approximation is used to calculate the dispersion relation 
for the oscillator energy $E_{i}$ and transferred momentum q: 
\begin{equation} 
E_{i}(q) = E_{i}(0) + \frac{\hbar^{2}q^{2}}{2m_{\text{e}}} 
\end{equation} 
The lower and upper limits of the transferred energy during an inelastic 
scattering event are as follows: 
\begin{equation} 
W_{-} = I_{\text{p}} 
\end{equation} 
\begin{equation} 
W_{+} = \frac{4Em_{1}m_{2}}{(m_{1}+m_{2})^{2}} 
\end{equation} 
where $E$ is the incident energy, and $m_{1}$ and $m_{2}$ are the masses of the 
two scattering particles. 
For the elastic scattering between electrons and atoms, i.e., the 
electron-phonon interaction, the Mott scattering cross section is employed with 
a modified Moli\`ere screening parameter $K_{\text{scr}}$,\cite{Jenkins2012} given 
by the following equation: 
\begin{equation} 
\sigma_{\text{elastic}} = \sigma_{\text{Mott}}K_{\text{scr}} 
\end{equation} 
The upper limit of transferred energy in the elastic scattering event is: 
\begin{equation} 
W_{+} = \mathrm{min}(\frac{4Em_{1}m_{2}}{(m_{1}+m_{2})^{2}}, 
\hbar\omega_{\text{Debye}}) 
\end{equation} 
where $\omega_{\text{Debye}}=(6\pi^{2}n_{\text{at}})^{\frac{1}{3}}v_{\text{s}}$ is the Debye 
frequency and $v_{\text{s}}$ is the speed of sound in \ga. 
 
Several additional parameters were incorporated into the TREKIS calculations: 
the speed of sound was set to $5,250~\mathrm{m/s}$,\cite{wright_acoustic_2025} the density 
of \ga\ to $5.88~\mathrm{g/cm}^3$, and the band gap to 
$4.9~\mathrm{eV}$.\cite{zhang_tight-binding_2022} 

\section{Technical details for MD simulations}
\label{app:md}
All MD simulations of SHI irradiation on $\beta$-\ga\ were performed using GPUMD. 
The simulations were initialized with a $\beta$-\ga\ lattice configuration oriented with the 
[010] crystallographic direction ($y$-axis) aligned along the ion penetration direction. 
Each simulation ran for a total of 200~ps in the NVE ensemble, with boundary temperature 
controlled using Nose-Hoover chain thermostats with coupling parameters tuned to 1,000. 
The time step was dynamically adjusted to mitigate instabilities arising from extreme 
temperature gradients: 0.25~fs for the initial 10~ps, 0.5~fs for the subsequent 90~ps, 
and 1~fs for the final 100~ps. 
To minimize boundary effects, the simulation cell was constructed with dimensions of 
38.8 $\times$ 10.5 $\times$ 39.5~nm, containing 1,457,920 atoms in total. 
To verify the sufficiency of the cell thickness, we also tested a model with dimensions of 
38.8 $\times$ 21.0 $\times$ 39.5~nm and found no significant difference in the final 
results between the standard and doubled-thickness models. 

\bibliography{main}

\end{document}


\title{Supplemental Material for \\ ``A Neuroevolution Potential for Gallium Oxide: \\
Accurate and Efficient Modeling of Polymorphism and Swift Heavy-Ion Irradiation''}

\author{Yaohui Gu}
\affiliation{State Key Laboratory of Heavy Ion Science and Technology, Institute of Modern Physics, Chinese Academy of Sciences, Lanzhou 730000, China}
\affiliation{School of Nuclear Science and Technology, University of Chinese Academy of Sciences, Beijing 100049, China}
\author{Binbo Li}
\thanks{Y.G. and B.L. contributed equally to this work.}
\affiliation{State Key Laboratory of Heavy Ion Science and Technology, Institute of Modern Physics, Chinese Academy of Sciences, Lanzhou 730000, China}
\affiliation{School of Nuclear Science and Technology, University of Chinese Academy of Sciences, Beijing 100049, China}
\author{Linyang Jiang}
\affiliation{State Key Laboratory of Heavy Ion Science and Technology, Institute of Modern Physics, Chinese Academy of Sciences, Lanzhou 730000, China}
\affiliation{School of Nuclear Science and Technology, University of Chinese Academy of Sciences, Beijing 100049, China}
\author{Yuhui Hu}
\affiliation{State Key Laboratory of Heavy Ion Science and Technology, Institute of Modern Physics, Chinese Academy of Sciences, Lanzhou 730000, China}
\affiliation{School of Nuclear Science and Technology, University of Chinese Academy of Sciences, Beijing 100049, China}
\affiliation{School of Materials $\&$ Energy, Lanzhou University, Lanzhou, 730000, China}
\author{Wenqiang Liu}
\affiliation{State Key Laboratory of Heavy Ion Science and Technology, Institute of Modern Physics, Chinese Academy of Sciences, Lanzhou 730000, China}
\affiliation{School of Nuclear Science and Technology, University of Chinese Academy of Sciences, Beijing 100049, China}
\author{Lijun Xu}
\affiliation{State Key Laboratory of Heavy Ion Science and Technology, Institute of Modern Physics, Chinese Academy of Sciences, Lanzhou 730000, China}
\affiliation{School of Nuclear Science and Technology, University of Chinese Academy of Sciences, Beijing 100049, China}
\author{Pengfei Zhai}
\affiliation{State Key Laboratory of Heavy Ion Science and Technology, Institute of Modern Physics, Chinese Academy of Sciences, Lanzhou 730000, China}
\affiliation{School of Nuclear Science and Technology, University of Chinese Academy of Sciences, Beijing 100049, China}
\author{Haizhou Xue}
\affiliation{State Key Laboratory of Heavy Ion Science and Technology, Institute of Modern Physics, Chinese Academy of Sciences, Lanzhou 730000, China}
\affiliation{School of Nuclear Science and Technology, University of Chinese Academy of Sciences, Beijing 100049, China}
\author{Jie Liu}
\affiliation{State Key Laboratory of Heavy Ion Science and Technology, Institute of Modern Physics, Chinese Academy of Sciences, Lanzhou 730000, China}
\affiliation{School of Nuclear Science and Technology, University of Chinese Academy of Sciences, Beijing 100049, China}
\author{Jinglai Duan}
\thanks{Corresponding author. Email: j.duan@impcas.ac.cn}
\affiliation{State Key Laboratory of Heavy Ion Science and Technology, Institute of Modern Physics, Chinese Academy of Sciences, Lanzhou 730000, China}
\affiliation{School of Nuclear Science and Technology, University of Chinese Academy of Sciences, Beijing 100049, China}
\affiliation{Advanced Energy Science and Technology Guangdong Laboratory, Huizhou 516000, China}

\date{\today}

\maketitle

\section{Extra Figures}

\begin{figure}[ht]
\centering
\includegraphics[width=0.8\linewidth]{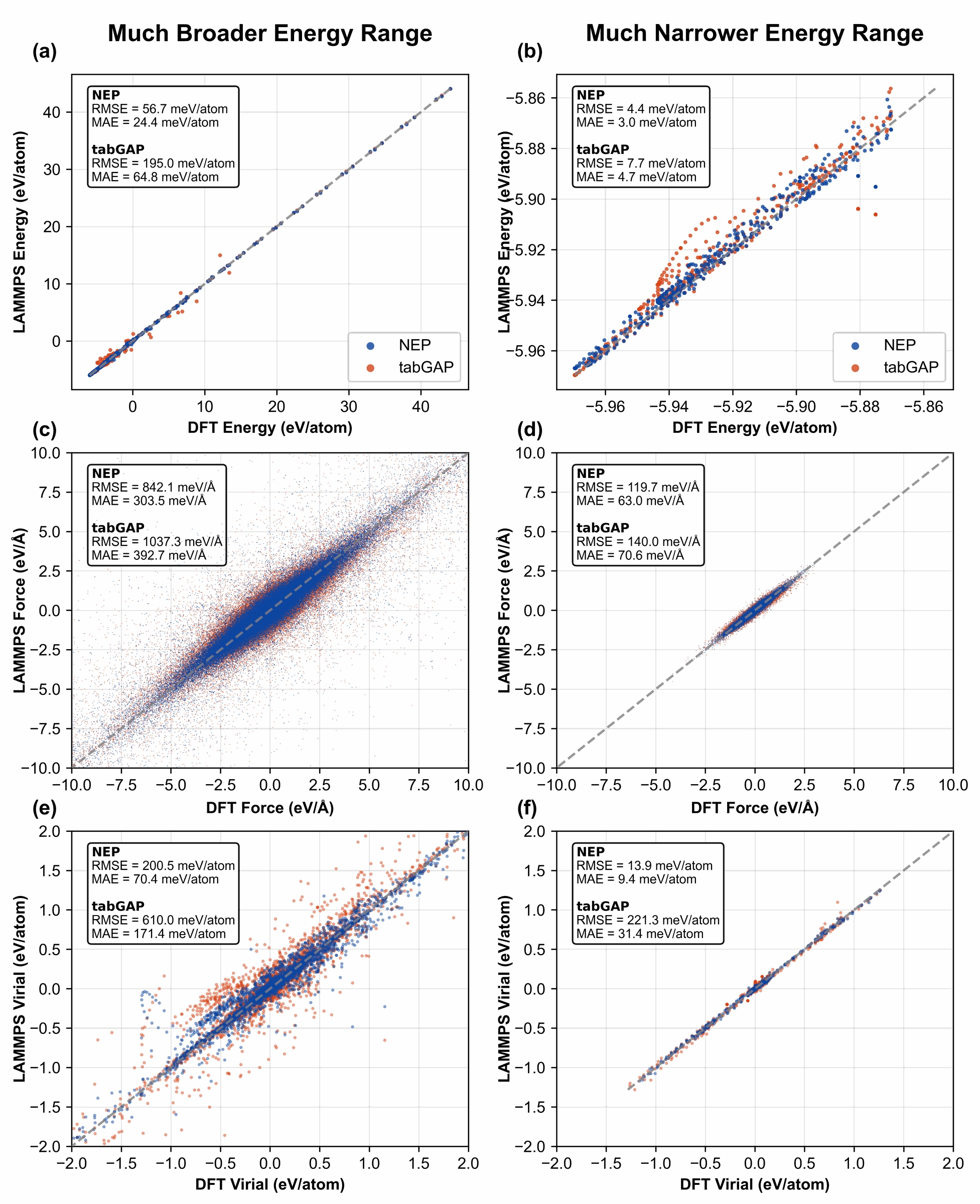}
\caption{Detailed comparison of the prediction accuracy of the NEP and tabGAP models.
Blue and red dots represent NEP and tabGAP results, respectively.
Panels (a), (c), and (e) show the predicted energies, forces,
and virials of the entire GAP-dataset configurations.
Panels (b), (d), and (f) present the corresponding results
for a much narrower energy window from $-5.9$ to $-5.8$~eV/atom.
\textit{Alt text:} Six scatter plots compare DFT reference values with model-predicted values for energy, force, and virial over the full GAP dataset and over a very narrow low-energy window. The full-dataset panels show much broader scatter, while the narrow-window panels show tightly grouped points near the diagonal. Across these comparisons, NEP generally remains closer to the diagonal than tabGAP, indicating better overall accuracy.}
\label{fig:S1}
\end{figure}

\begin{figure}[ht]
\centering
\includegraphics[width=\linewidth]{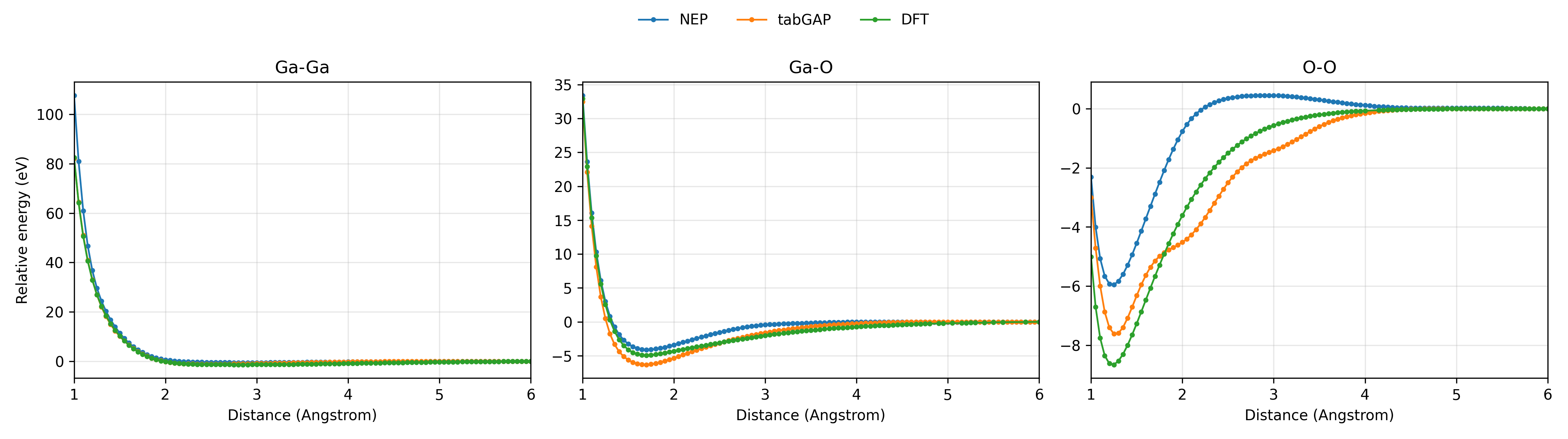}
\caption{Pair-energy curves for Ga--Ga, Ga--O, and O--O dimers from 1 to 6~\AA\ on a linear energy scale.
NEP, tabGAP, and DFT results are shown in blue, orange, and green, respectively.
The curves provide a direct comparison of isolated dimer interactions from the chemically relevant distance range to the far-field region.
\textit{Alt text:} Three line plots show relative dimer energy as a function of distance from 1 to 6~\AA\ for Ga--Ga, Ga--O, and O--O pairs. Overall, NEP and tabGAP give broadly comparable pair-energy behavior on this linear scale, with similar energy magnitudes and long-range convergence toward zero, although their agreement with DFT varies by pair type. For Ga--Ga, NEP, tabGAP, and DFT nearly overlap, with a steep repulsive wall near 1~\AA\ that rapidly decays to almost zero beyond about 2~\AA. For Ga--O, all three methods show short-range repulsion followed by an attractive well, but NEP gives a shallower minimum than tabGAP and DFT before all curves approach zero at large distance. For O--O, the disagreement is much larger: DFT has the deepest attractive well, tabGAP is somewhat shallower, and NEP shows a shallower minimum followed by a positive hump at intermediate distances before converging toward zero.}
\label{fig:S2}
\end{figure}

\begin{figure}[ht]
\centering
\includegraphics[width=\linewidth]{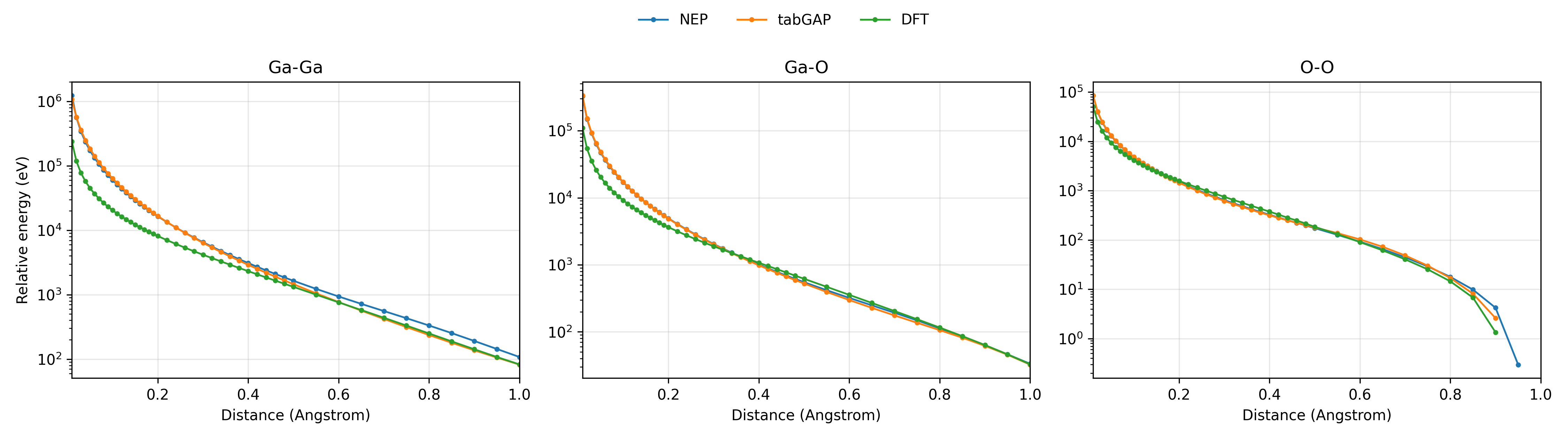}
\caption{Short-range pair-energy curves for Ga--Ga, Ga--O, and O--O dimers in the highly repulsive regime.
The energies are plotted on a logarithmic scale to compare NEP, tabGAP, and DFT from sub-angstrom separations up to 1~\AA.
This test verifies that NEP does not exhibit an obvious disadvantage relative to tabGAP in the short-range repulsive region relevant to high-energy collision processes.
\textit{Alt text:} Three logarithmic line plots compare short-distance dimer energies for Ga--Ga, Ga--O, and O--O pairs from sub-angstrom separations to 1~\AA. For all pairs, the repulsive energy increases rapidly as the atoms approach each other. In the O--O panel, NEP, tabGAP, and DFT agree very well across almost the entire distance range. For the Ga-containing Ga--Ga and Ga--O pairs, NEP and tabGAP also agree closely with each other, while the DFT curve lies lower at the shortest separations; this difference may arise from the use of non-all-electron pseudopotentials chosen to keep the benchmark calculation settings consistent.}
\label{fig:S3}
\end{figure}

\begin{figure}[ht]
\centering
\includegraphics[width=0.8\linewidth]{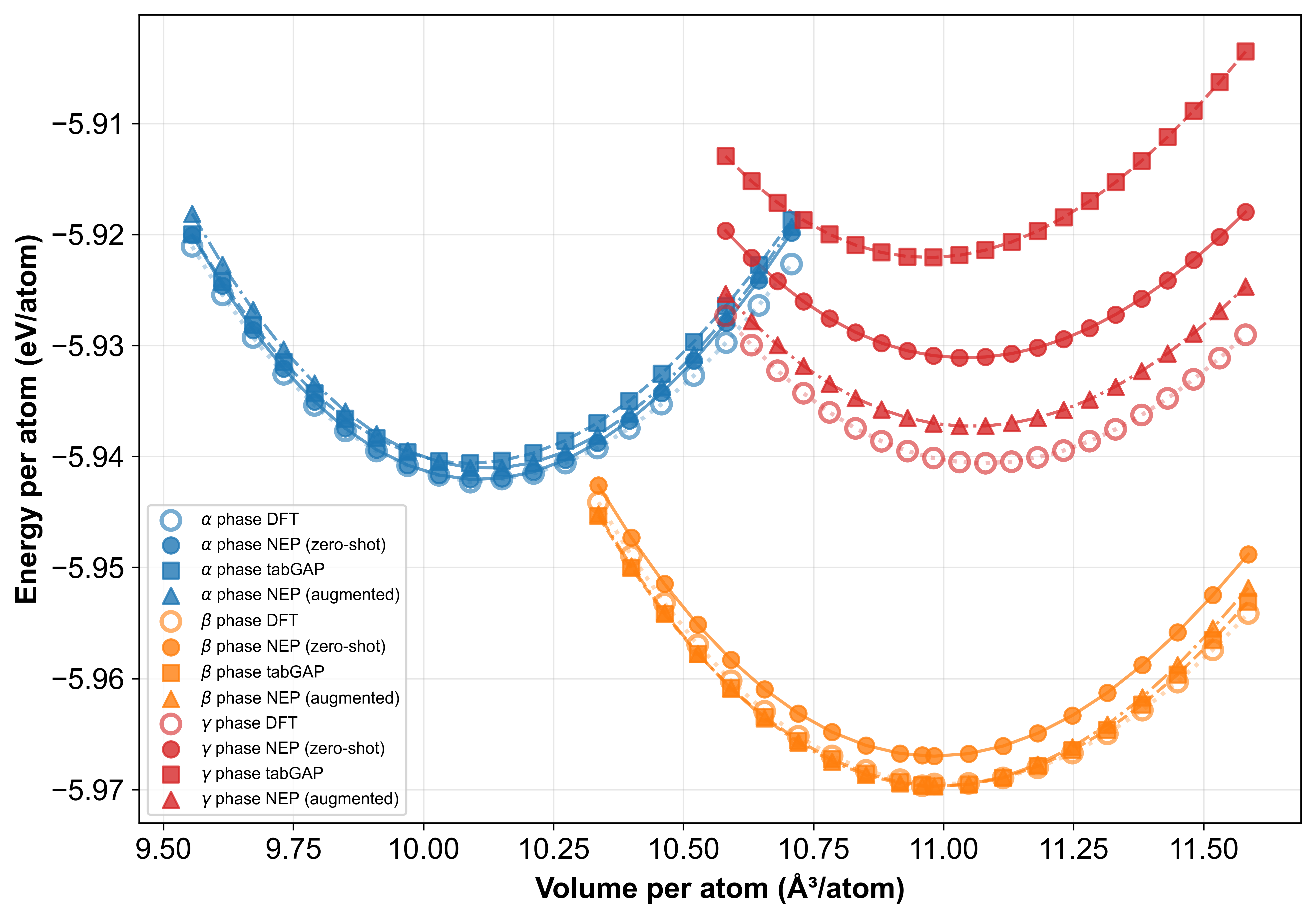}
\caption{Energy-volume curves for the $\alpha$, $\beta$ and $\gamma$ phases 
predicted by zero-shot NEP (solid circles), NEP with augmented training set (solid triangles), 
and tabGAP (solid squares), compared with DFT reference 
data (open circles). Different phases are color-coded as indicated.
\textit{Alt text:} A multi-curve plot shows energy per atom versus volume per atom for the alpha, beta, and gamma phases. For each phase, DFT reference data are compared with zero-shot NEP, augmented-data NEP, and tabGAP predictions. The augmented NEP curves lie closest to DFT, especially for the gamma phase, while zero-shot NEP and tabGAP deviate more strongly. The figure emphasizes that adding targeted training data improves the description of the gamma-phase equation of state.}
\label{fig:S4}
\end{figure}

\begin{figure}[ht]
\centering
\includegraphics[width=0.8\linewidth]{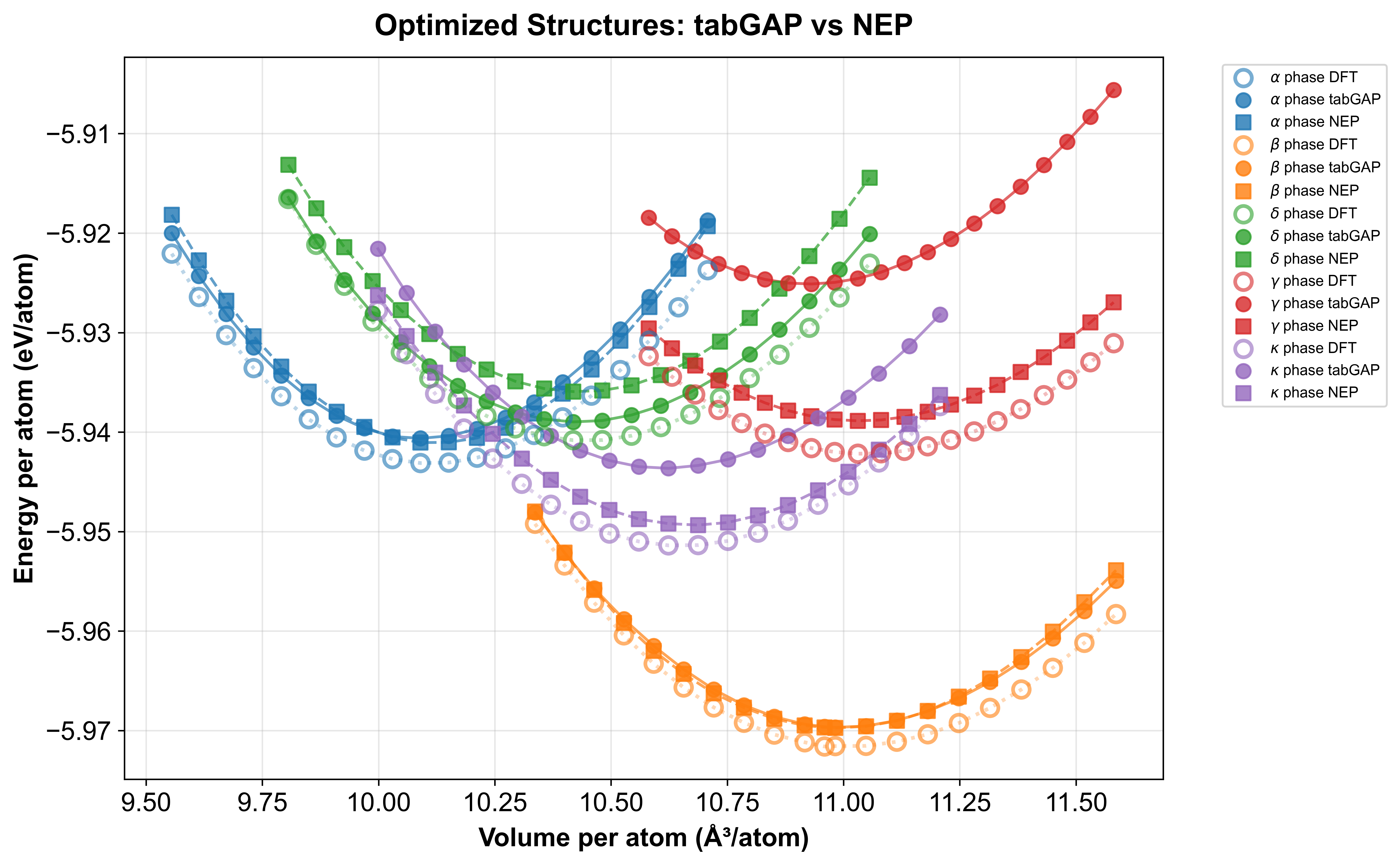}
\caption{Energy-volume curves for geometry-optimized $\alpha$, $\beta$, $\delta$, $\gamma$, and $\kappa$ phase Ga$_2$O$_3$ configurations.
The structures used in the corresponding energy-volume comparison were optimized with VASP, followed by single-point energy calculations and predictions using tabGAP and NEP.
Open circles denote DFT reference energies, filled circles denote tabGAP predictions, and filled squares denote NEP predictions.
The optimized-structure results lead to the same conclusions as the original energy-volume comparison, indicating that geometry optimization does not affect the main conclusions and further validating the reliability of the VASP calculation protocol.
\textit{Alt text:} A multi-curve energy-volume plot compares optimized alpha, beta, delta, gamma, and kappa phase Ga$_2$O$_3$ structures. Each phase is shown with DFT open circles, tabGAP filled circles, and NEP filled squares. The tabGAP and NEP curves generally follow the DFT energy-volume trends after geometry optimization, and the relative behavior among the phases remains consistent with the original comparison, showing that geometry optimization does not change the main conclusions.}
\label{fig:S5}
\end{figure}

\begin{figure}[ht]
\centering
\includegraphics[width=0.8\linewidth]{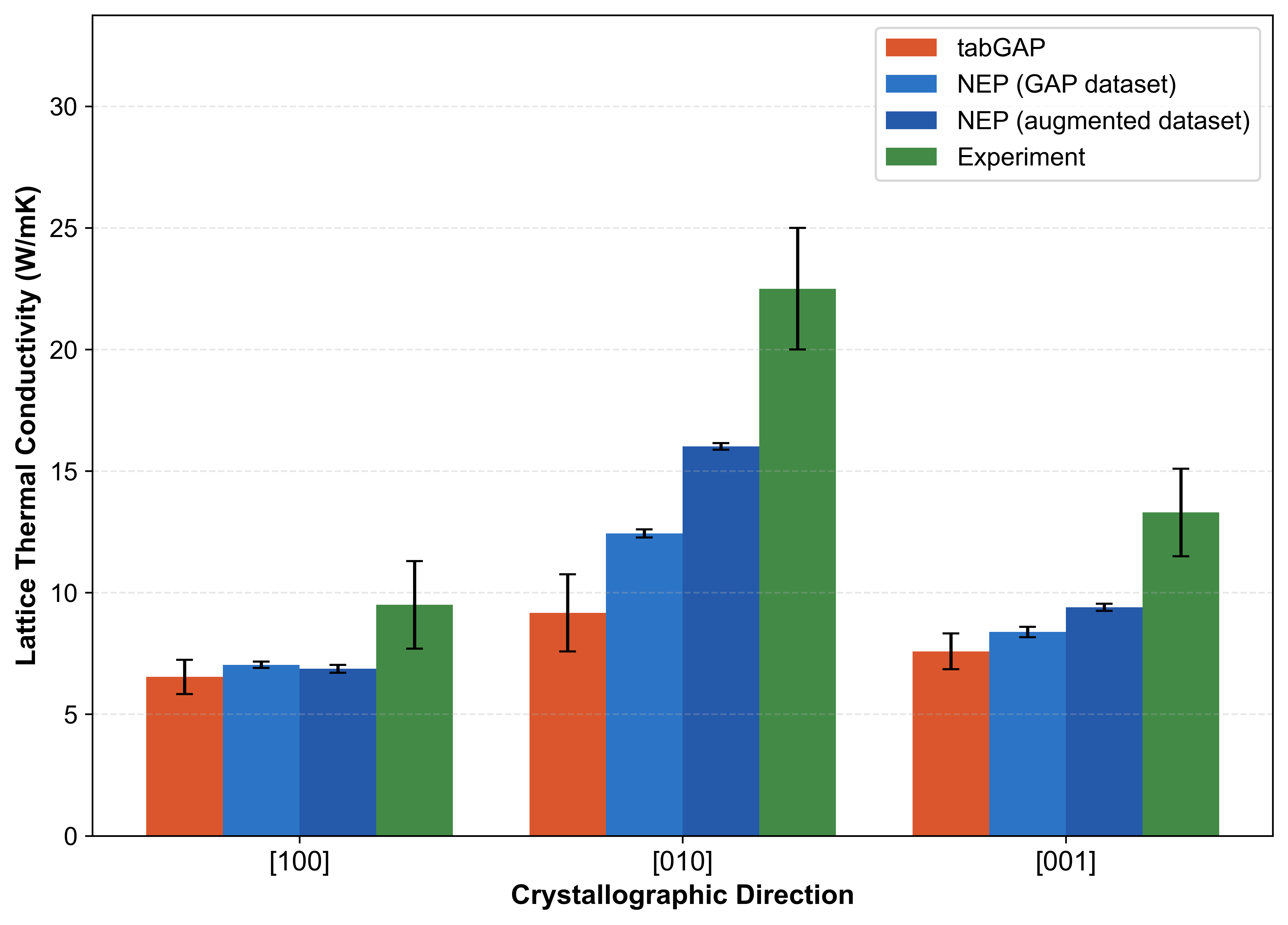}
\caption{Comparison of lattice thermal conductivity (LTC) predicted by the NEP and tabGAP models against experimental data.
The tabGAP predictions are obtained using the equilibrium molecular dynamics (EMD) method\cite{kuboStatisticalmechanicalTheoryIrreversible1957},
while the NEP predictions employ the more computationally efficient homogeneous nonequilibrium molecular dynamics (HNEMD) method\cite{fanHomogeneousNonequilibriumMolecular2019}.
\textit{Alt text:} A grouped bar chart compares lattice thermal conductivity along the [100], [010], and [001] crystallographic directions for tabGAP, NEP trained on the GAP dataset, NEP trained on the augmented dataset, and experiment. In all series, the [010] direction has the highest conductivity. The augmented NEP model reproduces this anisotropy more closely than the other simulations, although all calculated values remain lower than experiment.}
\label{fig:S6}
\end{figure}

\begin{figure}[ht]
\centering
\includegraphics[width=0.8\linewidth]{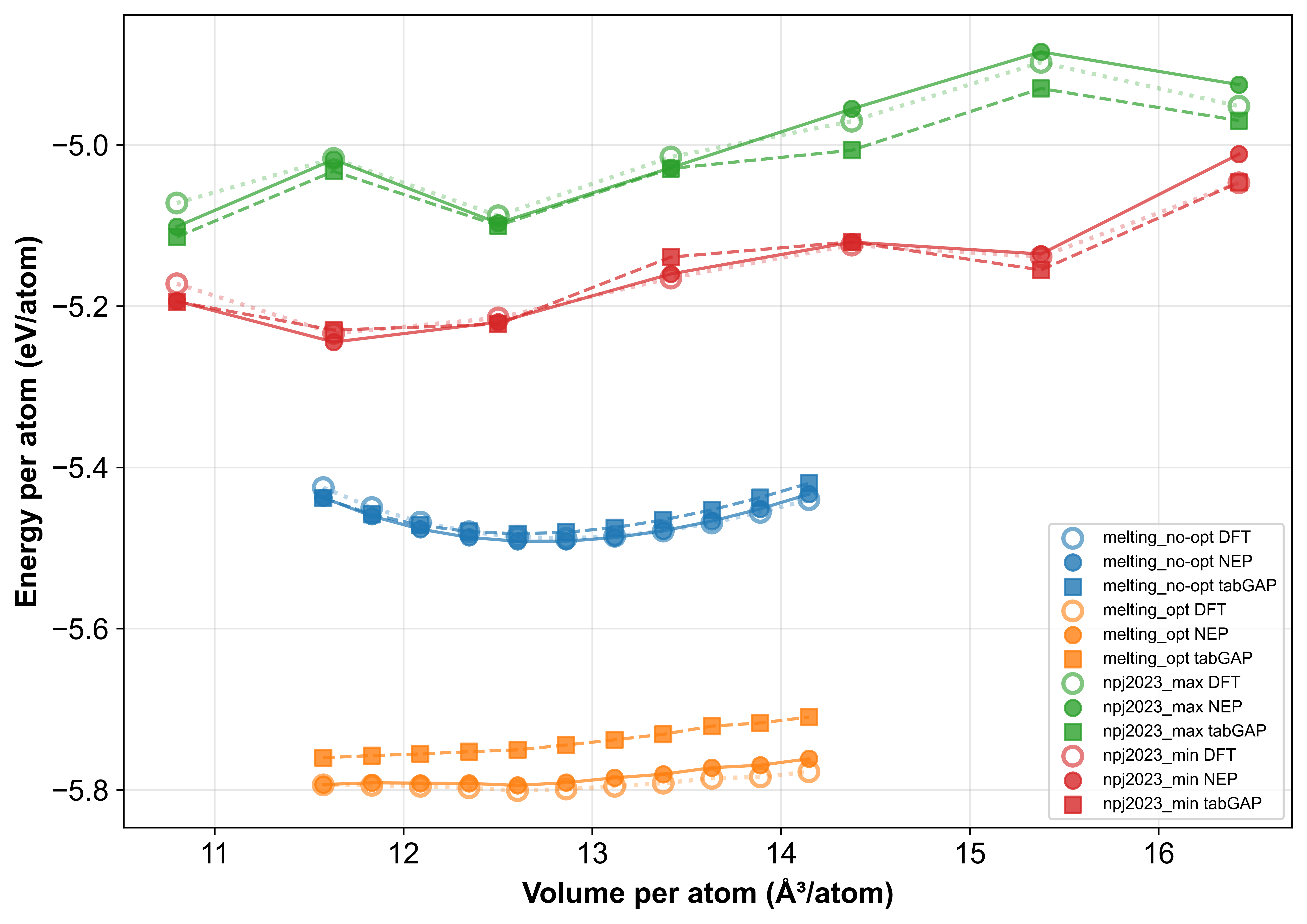}
\caption{Energy-volume comparison for noncrystalline Ga$_2$O$_3$ configurations.
Curves labeled \texttt{npj2023\_min} and \texttt{npj2023\_max} correspond to structures from the training dataset of Zhao \textit{et al.}\cite{zhaoComplexGa2O3Polymorphs2023}.
Curves labeled \texttt{melting\_no-opt} and \texttt{melting\_opt} correspond to additional molten-state configurations prepared in this work by stretching or compressing molten structures without and with subsequent geometry optimization, respectively.
Open circles denote DFT reference energies, filled circles denote NEP predictions, and filled squares denote tabGAP predictions.
\textit{Alt text:} A multi-curve plot compares energy per atom versus volume per atom for noncrystalline Ga$_2$O$_3$ configurations. The \texttt{melting\_no-opt} and \texttt{melting\_opt} curves are based on 2200~K molten structures generated in this work, and their energies lie much closer to the $\beta$-phase reference energy of about $-5.9$~eV/atom than the \texttt{npj2023\_min} and \texttt{npj2023\_max} curves from the Zhao et al. dataset, especially after geometry optimization for the orange \texttt{melting\_opt} curve. The comparison shows that NEP follows the DFT reference data more closely for configurations nearer to the ground-state energy, suggesting better performance for molten-state structures closer to the melting regime.}
\label{fig:S7}
\end{figure}

\begin{figure}[ht]
\centering
\includegraphics[width=0.55\linewidth]{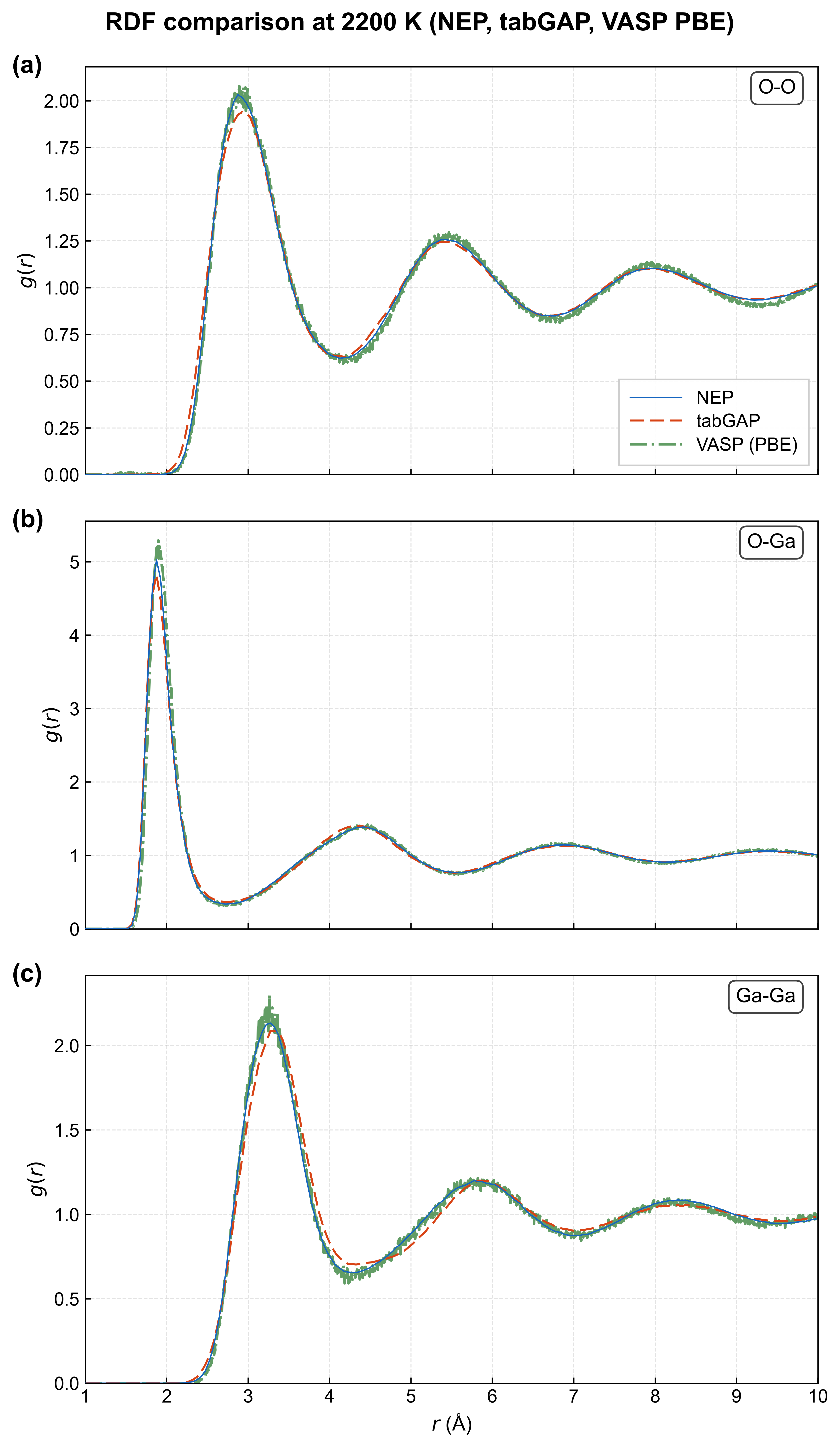}
\caption{Radial distribution functions (RDFs) of noncrystalline Ga$_2$O$_3$ at 2200~K and a density of $4.84~\mathrm{g}\cdot\mathrm{cm}^{-3}$.
Panels (a), (b), and (c) show the O--O, O--Ga, and Ga--Ga correlations, respectively.
NEP and tabGAP predictions are compared with VASP-PBE reference data.
The RDFs show that NEP follows the VASP-PBE reference more closely than tabGAP, particularly for the first-neighbor peak positions and local structural features.
\textit{Alt text:} Three vertically stacked RDF plots compare O--O, O--Ga, and Ga--Ga pair correlations at 2200 K. Blue NEP, orange tabGAP, and green VASP-PBE curves are shown in each panel. The NEP curves closely track the VASP-PBE peak positions and peak shapes, while tabGAP shows somewhat larger deviations in several first-neighbor features.}
\label{fig:S8}
\end{figure}

\begin{figure}[ht]
\centering
\includegraphics[width=0.65\linewidth]{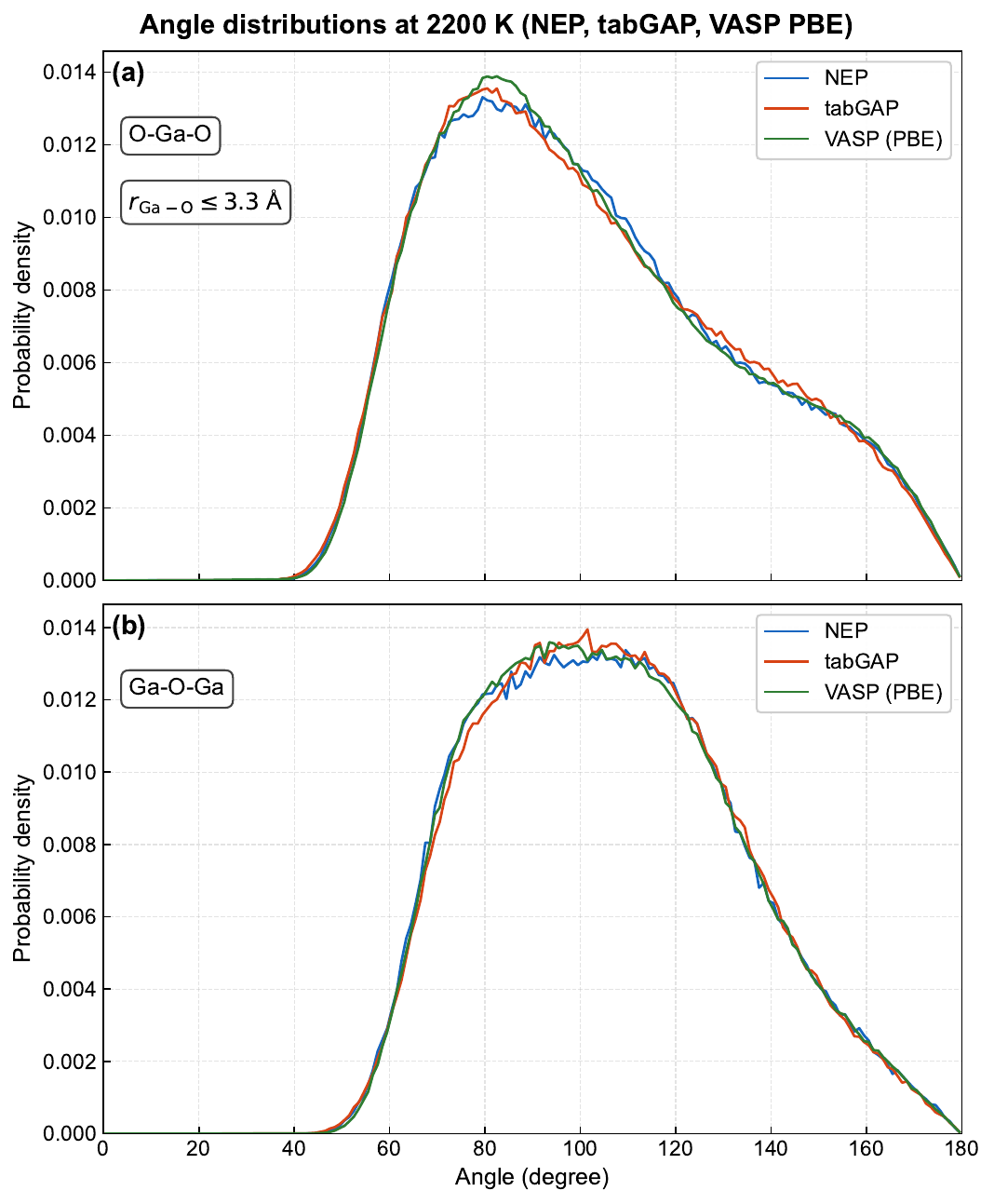}
\caption{Bond-angle distributions of noncrystalline Ga$_2$O$_3$ at 2200~K and a density of $4.84~\mathrm{g}\cdot\mathrm{cm}^{-3}$.
Panels (a) and (b) show the O--Ga--O and Ga--O--Ga angle distributions, respectively, using Ga--O pairs with $r_{\mathrm{Ga-O}} \leq 3.3$~\AA.
NEP and tabGAP both reproduce the VASP-PBE reference distributions reasonably well.
\textit{Alt text:} Two probability-density plots compare O--Ga--O and Ga--O--Ga angle distributions. The NEP, tabGAP, and VASP-PBE curves nearly overlap over most of the angular range, with broad maxima around intermediate angles and smooth decay toward 180 degrees.}
\label{fig:S9}
\end{figure}

\begin{figure}[ht]
\centering
\includegraphics[width=0.75\linewidth]{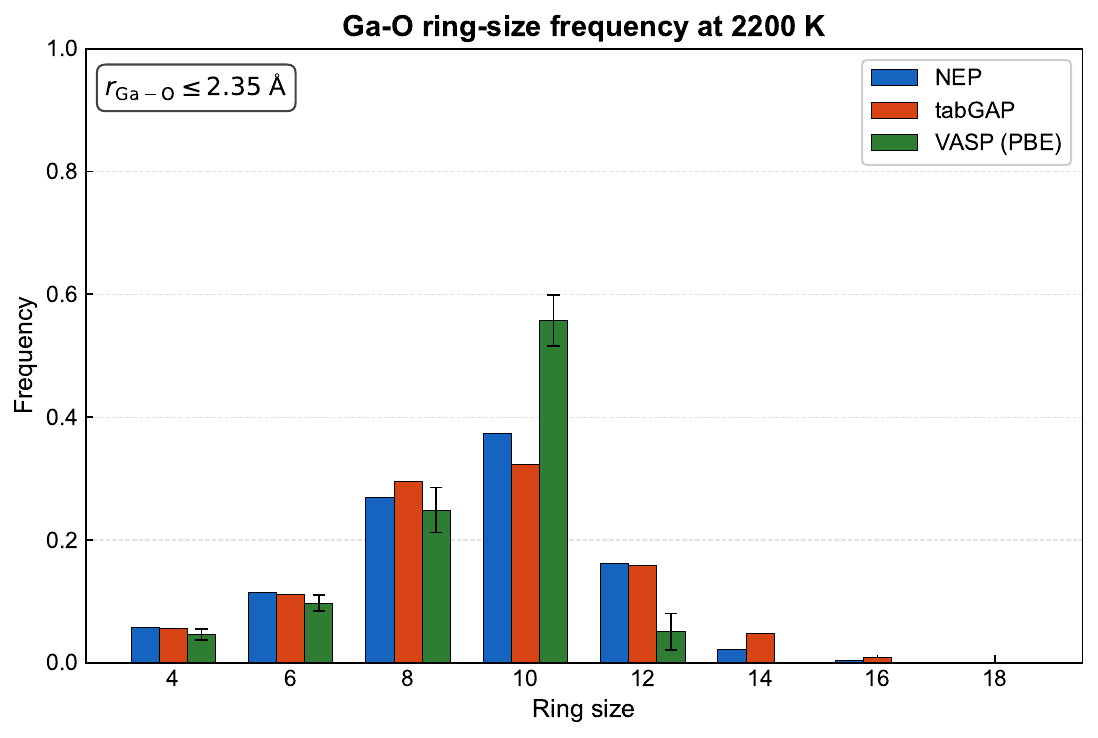}
\caption{Ga--O ring-size distributions of noncrystalline Ga$_2$O$_3$ at 2200~K and a density of $4.84~\mathrm{g}\cdot\mathrm{cm}^{-3}$.
Ring statistics are evaluated using Ga--O pairs with $r_{\mathrm{Ga-O}} \leq 2.35$~\AA.
The NEP and tabGAP distributions are broadly comparable, while VASP-PBE shows a stronger contribution from 10-membered rings under the same analysis criterion.
The error bars for the VASP-PBE data mainly reflect the statistical uncertainty associated with AIMD sampling in a relatively small simulation cell; the differences between VASP-PBE and the MLIP results may also be partly affected by this limited AIMD cell size.
\textit{Alt text:} A grouped bar chart compares Ga--O ring-size frequencies for NEP, tabGAP, and VASP-PBE. Most rings have sizes from 6 to 12, with all methods showing a dominant contribution around 8- and 10-membered rings. VASP-PBE has the largest 10-membered-ring frequency and includes visible error bars, which arise mainly from AIMD sampling in a relatively small simulation cell. NEP and tabGAP spread the distribution more evenly across neighboring ring sizes, and part of their difference from the VASP-PBE result may be related to the limited AIMD cell size rather than only to the model itself.}
\label{fig:S10}
\end{figure}

\begin{figure}[ht]
\centering
\includegraphics[width=0.8\linewidth]{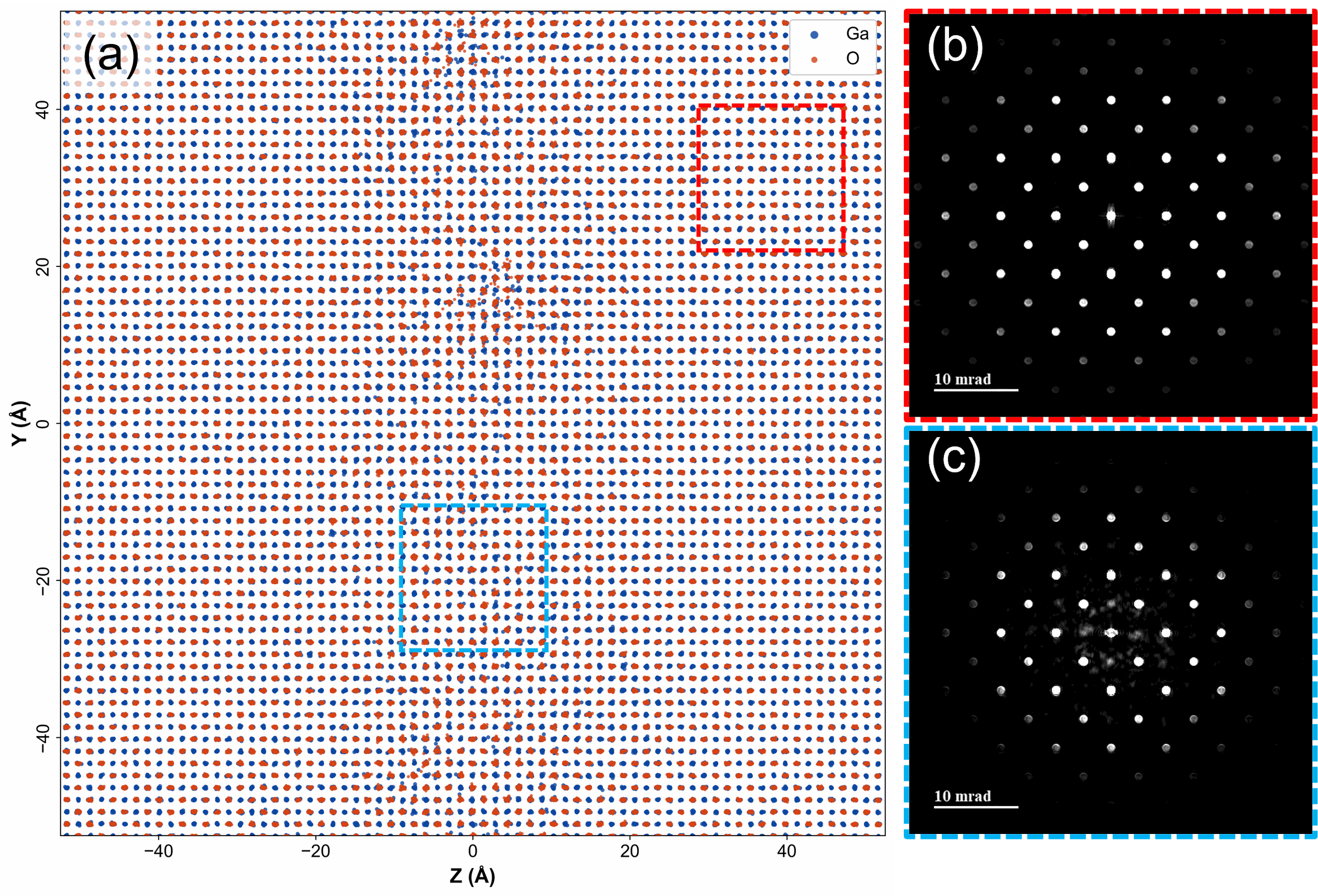}
\caption{(a) Longitudinal cross-sectional view of the SHI-irradiated $\beta$-Ga$_2$O$_3$ 
atomic structure with $S_e = 19.18$~keV/nm, the vertical direction is along [201] direction. 
(b) Simulated diffraction pattern for the pristine $\beta$ phase.
(c) Simulated diffraction pattern for the $\gamma$ phase.
\textit{Alt text:} Panel (a) shows a longitudinal atomic cross section of an irradiated beta-Ga$_2$O$_3$ track at an electronic energy loss of 19.18~keV/nm, with dashed boxes marking representative regions. Panels (b) and (c) show simulated diffraction patterns from selected areas. The pristine beta phase produces sharp, well-defined diffraction spots, whereas the gamma-phase region shows new characteristic diffraction spots associated with the gamma-phase structure. The figure demonstrates how irradiation-induced phase transformation from beta to gamma manifests in the diffraction pattern.}
\label{fig:S11}
\end{figure}

\begin{figure}[ht]
\centering
\includegraphics[width=\linewidth]{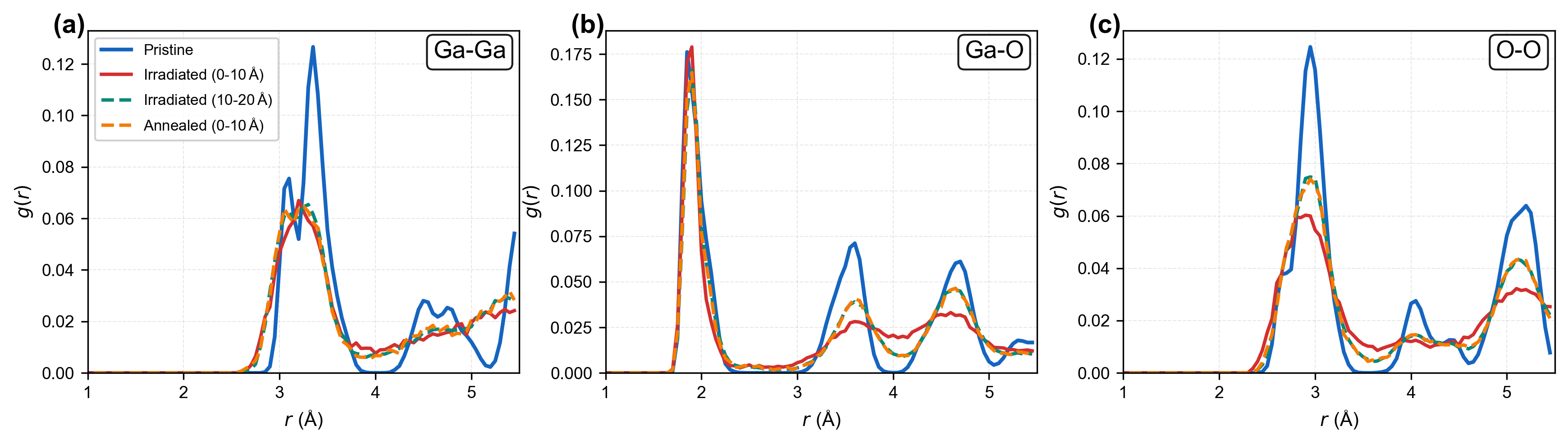}
\caption{Radial distribution functions (RDFs) of the ion track 
at $S_e = 22.16$~keV/nm before and after the annealing simulation at 800~K for 2~ns (Multimedia available online).
\textit{Alt text:} Three RDF plots compare Ga--Ga, Ga--O, and O--O pair correlations for pristine material, core and shell of the irradiated track, and the annealed core. Irradiation broadens and weakens peaks, showing increased disorder. After annealing at 800~K for 2~ns, several peaks within the core region become sharper and move closer to the shell (gamma) curves, indicating partial structural recovery and reorganization from the amorphous state to the gamma phase.}
\label{fig:S12}
\end{figure}

\begin{figure}[ht]
\centering
\includegraphics[width=0.6\linewidth]{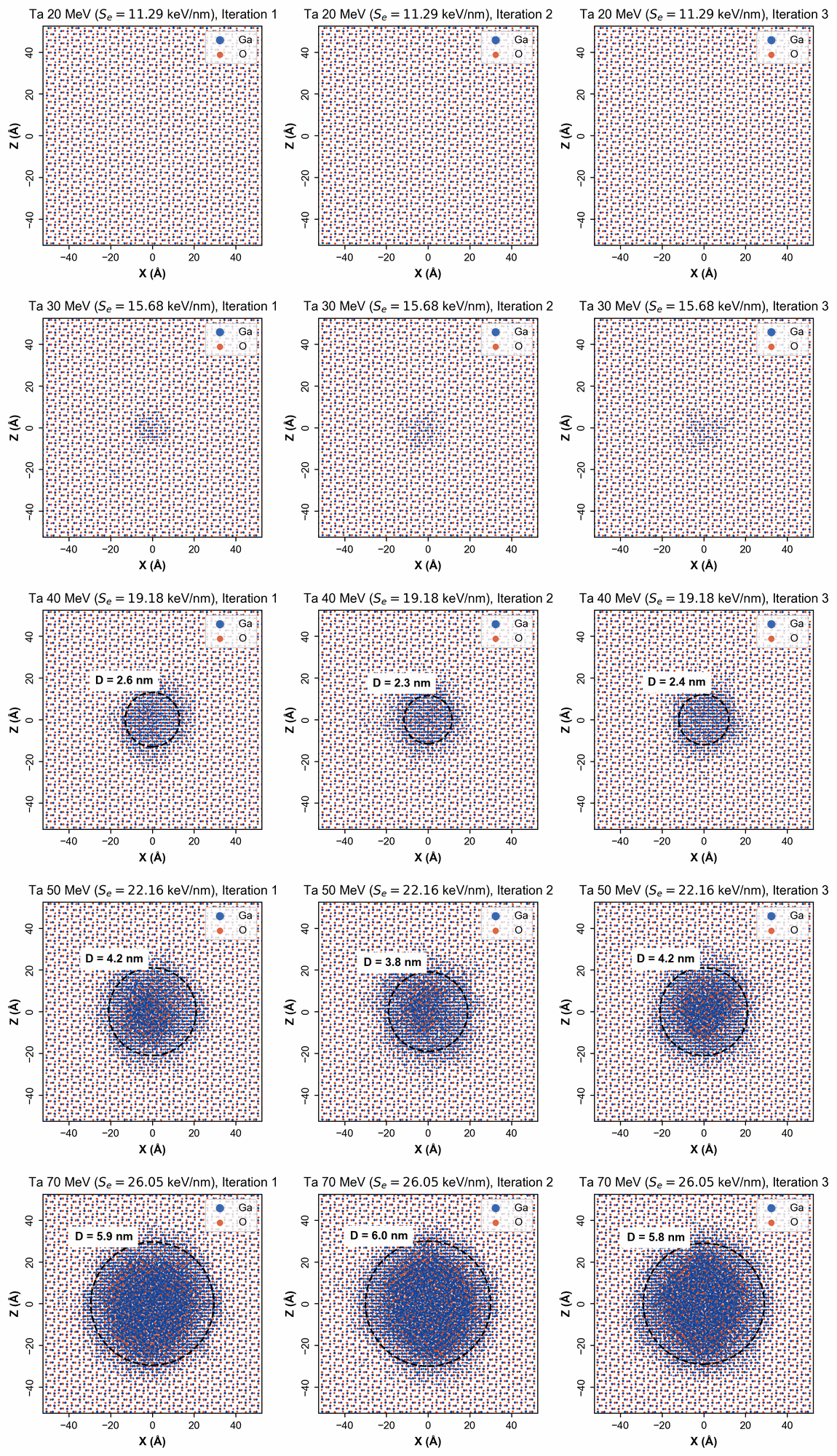}
\caption{Transverse cross-sectional views of ion tracks in $\beta$-Ga$_2$O$_3$ 
for electronic energy losses ranging from $S_e = 11.29$~keV/nm to 
$S_e = 26.05$~keV/nm. The measured track diameters are indicated for each case.
\textit{Alt text:} A matrix of transverse atomic cross sections shows three repeated simulations for each of five electronic energy losses between 11.29 and 26.05~keV/nm. At 11.29~keV/nm no visible track forms, and at 15.68~keV/nm only a faint damaged region appears. At 19.18, 22.16, and 26.05~keV/nm, clear tracks are present and their labeled diameters increase from about 2 to about 6~nm, although some run-to-run variation is visible.}
\label{fig:S13}
\end{figure}

\begin{figure}[ht]
\centering
\includegraphics[width=0.6\linewidth]{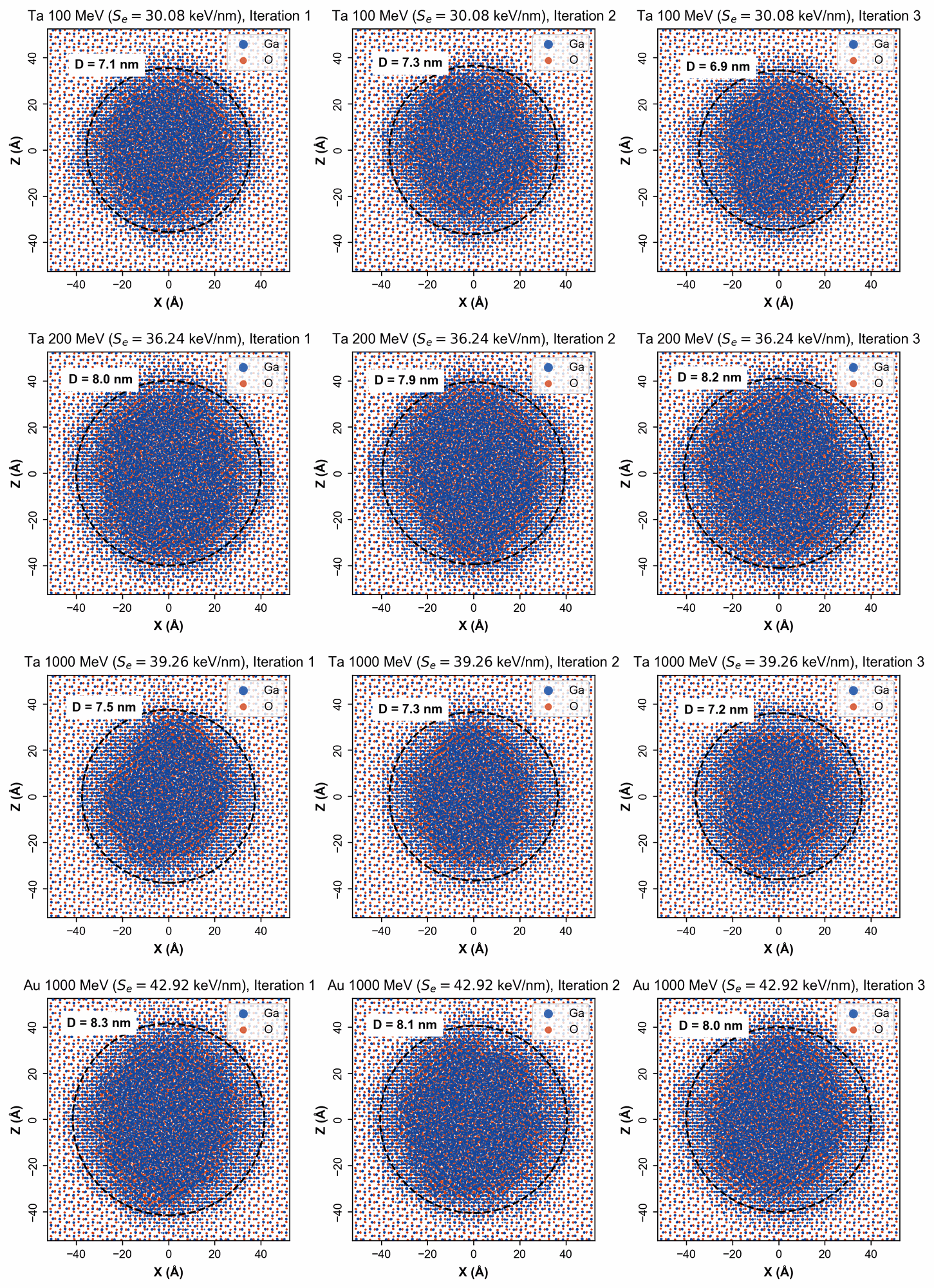}
\caption{Transverse cross-sectional views of ion tracks in $\beta$-Ga$_2$O$_3$ 
for electronic energy losses ranging from $S_e = 30.08$~keV/nm to 
$S_e = 42.92$~keV/nm. The measured track diameters are indicated for each case.
\textit{Alt text:} A matrix of transverse atomic cross sections shows three repeated simulations for each of four higher electronic energy losses between 30.08 and 42.92~keV/nm. All cases display large, well-developed tracks with labeled diameters of roughly 7 to 8.3~nm. Differences among repeats are modest, and the track size appears to approach saturation at the highest energy losses.}
\label{fig:S14}
\end{figure}

\begin{figure}[ht]
\centering
\includegraphics[width=0.8\linewidth]{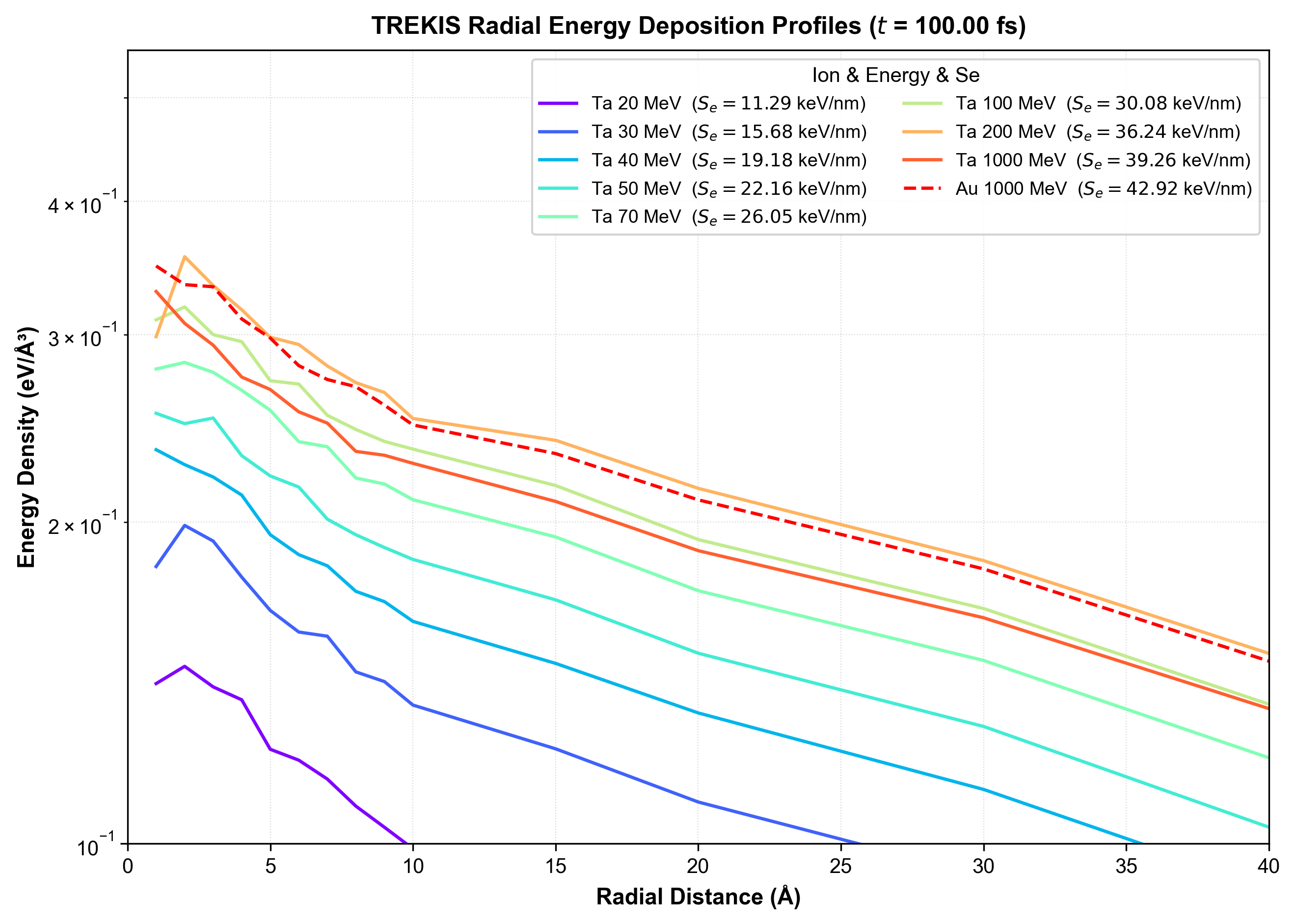}
\caption{Radial energy deposition profiles calculated by TREKIS for ions of different species and energies.
\textit{Alt text:} A line chart plots radial energy density versus radial distance for several ions and incident energies. All curves decrease with increasing distance from the track center, showing that deposited energy is most concentrated near the center. Higher-energy-loss cases lie higher overall and extend to larger radii, while some high-energy Ta and Au cases are close to each other near the center, indicating similar central energy densities despite different stopping powers.}
\label{fig:S15}
\end{figure}

\clearpage
\bibliography{sm}